\documentclass[aps,prl,twocolumn,superscriptaddress,groupedaddress]{revtex4}
\usepackage{subfig}
\usepackage{graphicx}  
\usepackage{dcolumn}   
\usepackage{bm}        
\usepackage{amssymb}   
\usepackage{slashed}
\usepackage{graphicx}				
\usepackage{amsmath}
\usepackage{mathtools}
\usepackage{tikz,pgf}
\usepackage{comment}
\usepackage{asymptote}
\usetikzlibrary{arrows,backgrounds}
\usetikzlibrary{fit,scopes,calc,matrix,positioning,decorations.pathmorphing}
\usepackage[all]{xy}
\usepackage{yfonts}
\newcommand{\bra}[1]{\ensuremath{\left\langle#1\right|}}
\newcommand{\ket}[1]{\ensuremath{\left|#1\right\rangle}}
\newcommand{\Bracket}[1]{\ensuremath{\left\langle#1\right\rangle}}

\DeclareFontFamily{OMS}{oasy}{\skewchar\font48 }
\DeclareFontShape{OMS}{oasy}{m}{n}{%
         <-5.5> oasy5     <5.5-6.5> oasy6
      <6.5-7.5> oasy7     <7.5-8.5> oasy8
      <8.5-9.5> oasy9     <9.5->  oasy10
      }{}
\DeclareFontShape{OMS}{oasy}{b}{n}{%
       <-6> oabsy5
      <6-8> oabsy7
      <8->  oabsy10
      }{}
\DeclareSymbolFont{oasy}{OMS}{oasy}{m}{n}
\SetSymbolFont{oasy}{bold}{OMS}{oasy}{b}{n}

\DeclareMathSymbol{\smallleftarrow}     {\mathrel}{oasy}{"20}
\DeclareMathSymbol{\smallrightarrow}    {\mathrel}{oasy}{"21}
\DeclareMathSymbol{\smallleftrightarrow}{\mathrel}{oasy}{"24}

\begin{document}
\title{Graviton mediated polarisation-polarisation entanglement of photons by means of the Schwinger Keldysh and Kadanoff Baym formalisms and Quantum Boltzmann equations}
\author{Andrei T. Patrascu}
\address{ELI-NP, Horia Hulubei National Institute for R\&D in Physics and Nuclear Engineering, 30 Reactorului St, Bucharest-Magurele, 077125, Romania\\
email: andrei.patrascu.11@alumni.ucl.ac.uk}
\begin{abstract}
In order to show that the graviton is a quantum entity an experiment is proposed that can show that quantum entanglement is produced by means of an exchange of gravitons. For this to be possible, one has to be able to witness the entanglement between the two objects considered in the experiment and to be able to eliminate other sources of entanglement like the exchange of virtual photons. Graviton mediated polarisation-polarisation photon entanglement is being analysed by methods originating from the Schwinger Keldysh and the Kadanoff Baym formalisms in conjunction with quantum Boltzmann type equations.  Applications in the context of cosmology are discussed.
\end{abstract}
\maketitle
\section{1. Introduction}
General relativity is one of the best tested theories to date and the most accurate theory of gravity. However, it is by construction a classical theory and because of that, it has some obvious limitations. General relativity doesn't describe the centre of a black hole or the origin of the universe, as it breaks down in both cases. As a classical theory, the procedure of quantisation can be performed in the standard way, and the results of this approach are meaningful as long as the regime is that of weak interactions and low energies. Therefore we can describe approximations of phenomena like the scattering of photons on gravitons or the interaction of other particles with gravitational waves. The problems related to quantum gravity appear when the energy of the processes increases enough to require higher order corrections. Equivalently, this means we are probing smaller and smaller regions of spacetime. Then, the problem of non-renormalisability comes in. All other interactions described by quantum field theories are renormalisable. Basically this means that the UV divergences emerging due to the improper treatment of small distance / high energy physics can be avoided by re-defining a finite number of parameters in the Lagrangian of our theory. This redefinition works in two stages. First we identify the divergences via a process called regularisation, then we absorb the divergent terms into unmeasurable, divergent corrections to our parameters. The result of the combination of the two divergences will result in a finite theory in which only a limited number of parameters had to be re-defined. 
With gravity, the problem cannot be solved in this way. If we start by following this path, we notice that in the process of absorbing the divergences, we always have to put in more divergent parameters than what we eliminate, leading to a theory that could only be rendered finite by an infinite number of corrective parameters. But a theory depending on an infinite number of parameters is hardly predictive. Therefore, the renormalisation procedure understood in the standard way fails to solve the divergence problems of quantum gravity. The solution to this problem was admirably found in string theory, where the small scale behaviour of the quantum field theory is replaced with the complex dynamics of a fundamental extended object, the string. The graviton emerges in string theory as a massless quantum mode quite naturally. As such it is a quantum object. However, there are alternative theories that consider gravity as a classical theory. To make sure that the graviton is indeed a quantum object, some experimental evidence is required. For that, it would be desirable to understand what it means for a particle to be quantum. Indeed, quantum mechanics changed our understanding of nature in some radical ways. One of the most important changes was in how we interpret probabilities of composite events. In classical physics, when we discuss the probability of an outcome at some end-point of a set of events, we calculate it as a product of probabilities for all the intermediate possible states over which we then sum. We can do this because we assume that any intermediate state has an underlying property that is actually realised, despite the fact that we are not capable of directly measuring it. This rather obvious and hidden assumption relies on the fact that for a certain classical observable, there must be one and only one outcome that is actually realised, the task of the experimenter being to simply measure it and obtain the "true value" for that property. Quantum mechanics drastically changed this viewpoint. Indeed, in quantum mechanics a property of a system may not even be determined without an experiment that could tell us something about the respective property. The property itself becomes well defined only if a certain measuring procedure for its outcomes is defined. In the absence of such a description, the property of the particle cannot be uniquely defined, and the various possibilities allowed by the laws of nature have to be taken into consideration as such. This observation leads to another interesting result, namely that if the evolution of the various possibilities is allowed in a standard unitary way, the interference of such evolutions leads to superpositions that in certain cases lead to different states of knowledge, that have a global meaning, without having a local meaning. To be more specific, let us consider two systems $A$ and $B$ with states in the respective spaces $H_A$ and $H_B$. When the two systems are being combined in an overarching system $AB$ the space of states describing the combined system, due to the nature of the evolution of the states of knowledge, will not only contain states corresponding to each state of $H_A$ paired with each state of $H_B$ but also states existing only in the tensor product of the two spaces $H_A \otimes H_B$. This space is far larger than $H_A \times H_B$ and the additional states found there are called entangled states. To produce such entangled states it is required to make use of some quantum intermediary, as it would be impossible to reach those entangled states by means of classical mediators. Quite obviously, if one allows for classical mediators then one assumes that all intermediate states must be perfectly defined for any observable, and hence there must be only one outcome for each intermediary point. This is called classical communication and will lead to an interaction in which the other system only receives classical information, hence no global irreducible state can be formed. However, if such an entangled state emerges, then the mediator must be a quantum entity. 
This argument can be used to show that the graviton is a quantum object by measuring how entanglement can be produced by means of gravitational interaction between two subsystems [1-4]. One can imagine several experimental difficulties associated to this task. It would be difficult to make sure the only interaction will be gravitational, and given the weak nature of gravitational interactions at the level of table-top experiments, the setup will clearly be challenging. 
In what follows I will show that in the context of quantum field theory, gravitational interaction with photons will result in an entangled state that can be associated with the exchange of gravitons. Indeed, it is known that if only local operations and classical communication exists one cannot produce entanglement between the two interacting subsystems. The resulting system will also be classical. In general, local operations and classical communication cannot increase the entanglement between the two subsystems. Therefore if the two subsystems turn out to be entangled, the mediator had to have quantum properties itself and the entanglement obtained must inherit properties of the mediator, for example, the entanglement must be decreasing with the distance between the two subsystems. 
To probe the entanglement between the resulting subsystems is no easy task. We will employ the quantum concurrence as a measure for the resulting entanglement and we will calculate it in the context of quantum field theory [5-9]. If the two systems have a property that is easily measured in order to detect entanglement, like, for example, polarisation, it would be even better. Therefore it may be interesting to consider the matter side of the problem as described by photons. 
This article is structured as follows: In section 2 the quantum nature of interaction is discussed as well as what it means for the mediator of an interaction to be quantum. In particular it is argued that a quantum mediator can entangle participants and that non-realised intermediate states do play a role in quantum mechanics, as not all intermediate states have uniquely defined properties. In section 3 a pedagogical introduction of the Schwinger Keldysh formalism is presented. Section 4 provides an introduction to the Kadanoff Baym formalism. Section 5 is dedicated to the introduction of the Boltzmann quantum equation. 
The actual calculations start in Section 6 where the model of photons scattering on gravitons is presented. 
Section 7 is dedicated to the photon-photon scattering mediated by gravitons. The use of the Quantum Boltzmann equation is presented in section 8 while the results related to the polarisation-polarisation entanglement are presented in section 9. Some conclusions and potential future applications are provided in section 10. 
\section{2. The quantum nature of interaction}
Following a quantum field theoretical description let us assume that the initial state of the matter system is separable, and let us define it as 
\begin{equation}
\ket{\psi_{i}}=\ket{0}_{A}\ket{0}_{B}
\end{equation}
We denote by $A$ and $B$ the states of the two subsystems that will interact via the exchange of our mediator. The states above are the ground states of the corresponding harmonic oscillators. An interaction can be seen as a perturbation which couples to the two states and can be characterised by a potential which we call $\lambda H_{AB}$. The system will be perturbed into a state that will be a linear combination of the excited harmonic oscillator states
\begin{equation}
\ket{\psi_{f}}=\frac{1}{\sqrt{N}}\sum_{n,N}C_{nN}\ket{n}\ket{N}
\end{equation}
Assuming the state is normalised we obtain $C_{00}=1$ and 
\begin{equation}
C_{nN}=\lambda\frac{\Bracket{nN|\hat{H}_{AB}|00}}{2E_{0}-E_{n}-E_{N}}
\end{equation} 
with $E_{0}$ the ground state energy of the harmonic oscillators and $E_{n}$ and $E_{N}$ the energies of the higher excited states. As stated in the introduction, the main difference between quantum and classical physics is that in quantum physics, one fundamentally has no one single outcome for a property of a system, therefore the property is not fundamentally defined through one single possible outcome, considered the "actual" value for that property. Instead, we have various possible intermediate outcomes that cannot be determined in the absence of the specification of a measuring procedure, and therefore, retain all their possible values as eigenvalues of an operator. Therefore in classical mechanics, the operator $\hat{H}$ would be simply a complex number, the actual result of the classical measurement of an intermediate state, and with that, due to the orthogonality of the many-particle states above, all the coefficients except for the vacuum one would be zero. The inability of defining an absolute intermediate state in the absence of a measurement prescription however makes it possible for our perturbed state to also contain many particle contributions. Note also that our operator had two indices $\hat{H}_{AB}$ as it acts on both systems $A$ and $B$. An operator acting only on one side would not be able to create entanglement. 
With the definitions above, it makes sense to re-write our final perturbed state in the form 
\begin{equation}
\begin{array}{c}
\ket{\psi_{f}} = (\ket{0}+\sum_{n>0}A_{n}\ket{n})\cdot (\ket{0}+\sum_{N>0}B_{N}\ket{N})+\\
\\
+\sum_{n,N>0}(C_{nN}-A_{n}B_{N})\ket{n}\ket{N}
\\
\end{array}
\end{equation}
with $A_{n}=C_{n0}$ and $B_{N}=C_{0N}$. This expression makes manifest the separable (first) part and the entangled (second) part of the final perturbed state. 
A measure for the quantitative level of entanglement between the two systems is the concurrence defined as 
\begin{equation}
C=\sqrt{2(1-tr[\hat{\rho}_{A}^{2}])}
\end{equation}
where
\begin{equation}
\hat{\rho}_{A}=\sum_{N}\Bracket{N|\psi_{f}}\Bracket{\psi_{f}|N}
\end{equation}
which becomes, using our perturbed final state,
\begin{equation}
\hat{\rho}_{A}=\frac{1}{N}\sum_{n,n',N}C_{nN}C^{*}_{n'N}\ket{n}\bra{n'}
\end{equation}
which leads to our measure for the strength of the entanglement
\begin{equation}
C=\sqrt{2(1-\frac{1}{N}\sum_{n,n',N,N'}C_{nN}C^{*}_{n'N}C_{n'N'}C^{*}_{nN'})}
\end{equation}

\section{3. The Schwinger-Keldysh formalism}
In order to describe the interaction between photons as mediated by gravitons in a context where we can follow the formation of entanglement as well as the dynamical evolution of the system and the decrease of entanglement, we have to consider systems far from equilibrium. In fact due to the extreme conditions where the gravitational interaction of photons would be significant, we have to focus in particular on such systems. It is expected that high photon density and hence intense photon beams would make a suitable environment for the distinction between gravitational photon coupling and the square diagram in which photons produce virtual electron-positron pairs. 
The result should appear as entangled narrow many particle modes with the entanglement decaying in time. This is particularly important for experimental verification of entanglement in high intensity laser facilities like the petawatt lasers in the CoReLS facility and others. Basically it means that instead of looking at entanglement between particles one could also look at entanglement and correlations between modes in multi-particle (identical particle) systems. This is basically the approach of choice when discussing for example gases of many identical particles, coherent photon beams, etc. In usual situations when entanglement between individual particles is considered one can define one-particle Hilbert spaces and analyse their product structure to identify entanglement. In a quantum computing context that implies manipulating single particles by means of various gates. However, this is not possible when one deals with, for example, ultracold gases, etc. A detailed discussion about this type of systems will be provided in a future article in the context of the Schwinger-Keldysh formalism and Boltzmann quantum equation. There is however a particularly useful discussion presented in ref. [31] 
 To follow such an entanglement dynamics a Schwinger-Keldysh approach is initially implemented, resulting in a far from equilibrium description amenable to analysis from a density matrix perspective. Discussions on the Schwinger-Keldysh formalism can be found in [22], [23], [24]. 
In general when we discuss equilibrium phenomena, the so called in-out formalism is being employed. This implies dividing the dynamical process into two sections, the remote past, and the remote future. Interactions are being turned on adiabatically at the remote past, the interaction process carries on in the intermediate stages, and the interactions are being turned off adiabatically in the remote future. This leads to the usual time-ordered correlation functions in which operators are being introduced in a fashion that preserves the time ordering of our interaction events, from an ground state defined in thermal equilibrium, towards another ground state defined similarly. The evolution due to the interaction is then described by an S-matrix of the form 
\begin{equation}
\hat{S}(t,t')=e^{i\hat{H}_{0}t}e^{-\hat{H}(t-t')}e^{-i\hat{H}_{0}t'}
\end{equation}
However, when we deal with out of equilibrium systems, we may know the starting state well enough, but we cannot ascribe a clearly defined (let alone equilibrium) final state. To circumvent this problem, the in-in formalism has been devised which takes advantage of the fact that the evolution contour can be deformed such that it advances in time to an arbitrary state and then returns to the initial well known state, without being terminated in an otherwise undefined state out of equilibrium. This solves our problem of calculating matrix elements at the expense of two phenomena: first, we will have to deal with out of time order correlators, namely correlators that are not ordered in time, but instead obey another ordering rule, say a circular ordering, and second, we will have to double our fields for each instant in time, introducing forward going fields, as well as backwards going ones. Of course, those fields won't be independent, but their distinctiveness has to be properly taken into account to begin with. 
Two approaches can be devised for this method: one would be generalising the usual approach to quantum field theory in the second quantisation by doubling the fields and considering the relations between the Green functions while analysing the operators by the usual Wick rules. The other, somehow more formal, but at the same time more powerful, implies generalising the Feynman path integral approach by doubling the fields and obtaining the perturbative expansion while considering the functional derivative with respect to the sources associated to both forward and backward going fields. Both methods provide reliable results, and both are valid and have pros and cons. 
Say we want to calculate an expectation value $\Bracket{Q(\tau)}$ where 
\begin{equation}
Q(\tau)=\mathcal{O}_{1}(\tau,x_1)...\mathcal{O}_{N}(\tau,x_N)
\end{equation}
where the operators $\mathcal{O}_{i}(\tau,x_{i})$ are well defined local operators. We can write them in terms of the fields of the Lagrangian and they will take the form 
\begin{equation}
Q(\tau)=\phi^{A_1}(\tau,x_{1})...\phi^{A_{N}}(\tau,x_{N})
\end{equation}
In the interaction picture, in the equilibrium formalism, we can write 
\begin{equation}
\Bracket{Q(\tau)}=\bra{\Omega}\bar{F}(\tau,\tau_0)Q_{I}(\tau)F(\tau,\tau_{0})\ket{\Omega}
\end{equation}
where 
\begin{equation}
F(\tau,\tau_{0})=T exp(-i\int_{\tau_0}^{\tau}d\tau_{1}H_{I}(\tau_{1}))
\end{equation}
and
\begin{equation}
\bar{F}(\tau,\tau_{0})=\bar{T} exp(-i\int_{\tau_0}^{\tau}d\tau_{1}H_{I}(\tau_{1}))
\end{equation}
where $T$ and $\bar{T}$ represent time and anti-time ordering and $H_{I}(\tau)$ is the interacting part of the Hamiltonian. 
Using the notation
\begin{equation}
\Bracket{Q}=\bra{\Omega}Q(\tau)\ket{\Omega}
\end{equation}
we insert the unit decomposition in front of all operator insertions as follows
\begin{equation}
\Bracket{Q}=\sum_{\alpha}\bra{\Omega}\ket{\alpha}\bra{\alpha}\phi^{A_{1}}(\tau,x_{1})...\phi^{A_{N}}(\tau,x_{N})\ket{\Omega}
\end{equation}
where 
\begin{equation}
\sum_{\alpha}\ket{\alpha}\bra{\alpha}=1
\end{equation}
on a time-slice $\Sigma_{f}$ for a time $\tau_f\geq \tau$. 
The matrix element $\bra{\Omega}\ket{\alpha}$ looks like the conjugation of a vacuum amplitude, but with the time order of the in and out states reversed. This would then be an anti-time-ordered factor. The other factor $\bra{\alpha}Q\ket{\Omega}$ is a time ordered factor. To obtain the path integral formulation we introduce a foliation between the initial slice $\Sigma_0$ and the final slice $\Sigma_f$ and insert the unity decomposed in the field operator and in the momentum eigenbasis on each of the slices in the foliation
\begin{equation}
1=\sum_{\phi(\tau_{i})}\ket{\phi(\tau_{i})}\bra{\phi(\tau_{i})}
\end{equation}
and 
\begin{equation}
1=\sum_{\pi(\tau_{i})}\ket{\pi(\tau_{i})}\bra{\pi(\tau_{i})}
\end{equation}
and we write the path integral representation of the time ordered factor as 
\begin{widetext}
\begin{equation}
\begin{array}{c}
\bra{\alpha}Q_{I}(\tau)\ket{\Omega}=\int\mathcal{D}\phi_{+}\mathcal{D}\pi_{+}exp[i\int_{\tau_{0}}^{\tau_{f}}d\tau d^{3}x (\pi_{+A}\phi_{+}^{'A}-\mathcal{H}[\pi_{+},\phi_{+}])]\\
\\
\times \phi_{+}^{A_1}(\tau,x_{1})...\phi_{+}^{A_N}(\tau,x_{N})\bra{\alpha}\ket{\phi_{+}(\tau_f)}\bra{\phi_{+}(\tau_{0})}\ket{\Omega}
\end{array}
\end{equation}
\end{widetext}
and the anti-time ordered factor can be written as a path integral over the field configurations $\phi_{-}^{A}(\tau,x)$ and conjugate momenta $\pi_{-A}(\tau,x)$

\begin{widetext}
\begin{equation}
\begin{array}{c}
\bra{\Omega}\ket{\alpha}=\int\mathcal{D}\phi_{-}\mathcal{D}\pi_{-}exp[-i\int_{\tau_{0}}^{\tau_{f}}d\tau d^{3}x(\pi_{-A}\phi_{-}^{'A}-\mathcal{H}[\pi_{-},\phi_{-}])]\times\\
\\
\bra{\phi_{-}(\tau_f)}\ket{\alpha}\bra{\Omega}\ket{\phi_{-}(\tau_{0})}
\end{array}
\end{equation}
\end{widetext}
The fields and the momenta introduced above have a time ordered and respectively an anti-time ordered component, encoded in their $\pm$ indices. To obtain the expectation value of our observable now, we combine the two field Lagrangians 
\begin{equation}
\begin{array}{c}
\Bracket{Q}=\int\mathcal{D}\phi_{+}\mathcal{D}\pi_{+}\mathcal{D}\phi_{-}\mathcal{D}\pi_{-}\phi_{+}^{A_{1}}(\tau,x_{1})...\phi_{+}^{A_{N}}(\tau,x_{N})\times \\
\\
\times exp[i\int_{\tau_{0}}^{\tau_{f}}d\tau d^{3}x(\pi_{+A}\phi_{+}^{'A}-\mathcal{H}[\pi_{+},\phi_{+}])]\\
\\
\times exp[-i\int_{\tau_{0}}^{\tau_{f}}d\tau d^{3}x(\pi_{-A}\phi_{-}^{'A}-\mathcal{H}[\pi_{-},\phi_{-}])]\\
\\
\times\Bracket{\Omega|\phi_{-}(\tau_{0})}\Bracket{\phi_{+}(\tau_{0})|\Omega}\prod_{A,x}\delta(\phi_{+}^{A}(\tau_{f},x)-\phi_{-}^{A}(\tau_{f},x))\\
\end{array}
\end{equation}
The path integral remains unconstrained at both ends of the time interval, which means one has to integrate over all possible states of the form $\ket{\phi_{-}(\tau_{0})}$ and $\bra{\phi_{+}(\tau_{0})}$. This leads to two path integrals, one directed forward in time, the other backwards, both being linked together at the future time limit $\tau_{f}$ through the condition $\phi_{+}^{A}(\tau_{f})=\phi_{-}^{A}(\tau_{f})$. If the theory has no higher order derivative couplings, the momentum integral can be carried out because the Hamiltonian is quadratic in momentum and produces after integration a Gaussian. This leads then to 
\begin{widetext}
\begin{equation}
\begin{array}{c}
\Bracket{Q}=\int\mathcal{D}\phi_{+}\mathcal{D}\phi_{-}\phi_{+}^{A_{1}}(\tau,x_{1})...\phi_{+}^{A_{N}}(\tau,x_{N})exp[i\int_{\tau_{0}}^{\tau_{f}}d\tau d^{3}x(\mathcal{L}_{cl}[\phi_{+}]-\mathcal{L}_{cl}[\phi_{-}])]\\
\\
\times\Bracket{\Omega|\phi_{-}(\tau_{0})}\Bracket{\phi_{+}(\tau_{0})|\Omega}\prod_{A,x}\delta(\phi_{+}^{A}(\tau_{f},x)-\phi_{-}(\tau_{f},x))
\end{array}
\end{equation}
\end{widetext}
the delta functional in the last line has the role of smoothly connecting the two branches of time and anti-time field operators. The last two inner products, namely $\Bracket{\Omega|\phi_{-}(\tau_{0})}$ and $\Bracket{\phi_{+}(\tau_{0})|\Omega}$ give us the right prescription of contour integration. Those inner products are basically vacuum wave functionals represented in the field basis. Therefore they obey the vacuum equation $b_{A}\ket{\Omega}$ written in the field basis, with $b_{A}$ being the annihilation operator. We assume that the interactions are switched off at the remote past, making the field $\phi^{A}$ free at $\tau_{0}$. With this, and considering the fields also effectively massless at the remote past we have mode functions of the form 
\begin{equation}
u_{A}(\tau_{0},k)=\frac{1}{\sqrt{2k}}e^{-i k \tau}
\end{equation}
With this, the annihilation operator has the form
\begin{widetext} 
\begin{equation}
b_{A}=-i\int d^{3}x[a^{2}(\tau)u_{A}^{*}(\tau,-k)\phi_{A}(\tau,x)-u_{A}^{*}(\tau,-k)\pi_{A}(\tau,x)]e^{-i kx}
\end{equation}
\end{widetext}
The momentum operator in the field basis is
\begin{equation}
\pi_{A}(\tau,x)=-i\frac{\delta}{\delta\phi^{A}(\tau,x)}
\end{equation}
Therefore our vacuum equation $b_{A}\ket{\Omega}=0$ in the field basis $\ket{\phi_{+}(\tau_{0})}$ becomes

\begin{equation}
\begin{array}{c}
0=\int d^{3}xe^{-ikx}[\frac{\delta}{\delta\phi_{+}^{A}(\tau_{0},x)}-\frac{i u_{A}^{*'}(\tau_{0},-k)}{u_{A}^{*}(\tau_{0},k)}\phi_{+A}(\tau_{0},x)]\Bracket{\phi_{+}(\tau_{0})|\Omega}=\\
\\
=\int d^{3}x e^{-ikx}[\frac{\delta}{\delta\phi_{+}^{A}(\tau_{0},x)}+k\phi_{+A}(\tau_{0},x)]\Bracket{\phi_{+}(\tau_{0})|\Omega}\\
\end{array}
\end{equation}

This equation is being solved by 
\begin{widetext}
\begin{equation}
\begin{array}{c}
\Bracket{\phi_{+}(\tau_{0})|\Omega}=\mathcal{N}exp[-\frac{1}{2}\int d^{3}x d^{3}y \mathcal{E}_{AB}(\tau_{0};x,y)\phi_{+}(\tau_{0},x)\phi_{+}(\tau_{0},y)]=\\
\\
=\mathcal{N}exp[-\frac{\epsilon}{2}\int_{\tau_{0}}^{\tau_{f}}d\tau\int d^{3}x d^{3}y\mathcal{E}_{AB}(\tau,x,y)\phi_{+}^{A}(\tau,x)\phi_{+}^{A}(\tau,x)\phi_{+}^{B}(\tau,y)e^{\epsilon \tau}]\\
\\
\end{array}
\end{equation}
\end{widetext}
We introduced a normalisation factor $\mathcal{N}$ and an infinitesimal parameter $\epsilon$ that will parametrise the deformation of the integration contour. The expression for $\mathcal{E}_{AB}(\tau;x,y)$ is obtained by introducing this assumed solution back into the vacuum state equation, the result being 
\begin{equation}
\mathcal{E}_{AB}(\tau;x,y)=\int \frac{d^{3}k}{(2\pi)^{3}}e^{ik(x-y)}k \delta_{AB}
\end{equation}
and therefore the solution to the vacuum equation will be
\begin{widetext}
\begin{equation}
\Bracket{\phi_{+}(\tau_{0})|\Omega}=\mathcal{N} exp[-\frac{\epsilon}{2}\int_{\tau_{0}}^{\tau_{f}}d\tau\int\frac{d^{3}k}{(2\pi)^{3}}k\phi_{+A}(\tau,k)\phi_{+}^{A}(\tau,-k)]
\end{equation}
\end{widetext}
neglecting higher order corrections in $\epsilon$.
For the other inner product we obtain similarly 

\begin{equation}
\Bracket{\Omega|\phi_{-}(\tau_{0})}=\mathcal{N}^{*}exp[-\frac{\epsilon}{2}\int_{\tau_{0}}^{\tau_{f}}d\tau\int\frac{d^{3}k}{(2\pi)^{3}}k\phi_{-A}(\tau,k)\phi_{-}^{A}(\tau,-k)]
\end{equation}

The normalisation factors are irrelevant because they cancel out with the calculation of the expectation values. 
If we substitute back into the classical Lagrangians we obtain 
\begin{equation}
\mathcal{L}[\phi_{\pm}]\rightarrow\mathcal{L}\phi_{\pm}]\pm \frac{i \epsilon}{2}\int d\tau\int\frac{d^{3}k}{(2\pi)^{3}}k\phi_{\pm A}(k)\phi_{\pm}^{A}(-k)
\end{equation}
The additional term depending on $\epsilon$ produces a shift in the path for the time ordered variable 
\begin{equation}
k\tau\rightarrow (1-i\epsilon)k\tau
\end{equation}
and an opposite shift for the anti-time-ordered variable 
\begin{equation}
k\tau\rightarrow (1+i\epsilon)k\tau
\end{equation}
The time direction therefore gets slightly deformed into the complex plane in the upper, respectively lower part the for time-ordered and anti-time-ordered components. 
The two inner products therefore provide the correct $i\epsilon$ prescription for the path integral. The expression for the expectation value therefore will be
\begin{widetext}
\begin{equation}
\Bracket{Q}=\int\mathcal{D}\phi_{+}\mathcal{D}\phi_{-} \phi_{+}^{A_{1}}(\tau,x_{1})...\phi_{+}^{A_{N}}(\tau,x_{N})exp[i\int_{\tau_{0}}^{\tau_{f}}d\tau d^{3}x(\mathcal{L}_{cl}[\phi_{+}]-\mathcal{L}_{cl}[\phi_{-}])]\prod_{A,x}\delta(\phi_{+}^{A}(\tau_{f},x)-\phi_{-}^{A}(\tau,x))
\end{equation}
\end{widetext}
Diagrammatic rules can be constructed in the standard fashion, by introducing sources for the two fields
\begin{widetext}
\begin{equation}
Z[J_{+},J_{-}]=\int\mathcal{D}\phi_{+}\mathcal{D}\phi_{-} exp[i\int_{\tau_{0}}^{\tau_{f}} d\tau d^{3}x (\mathcal{L}_{cl}[\phi_{+}]-\mathcal{L}_{cl}[\phi_{-}]+J_{+}\phi_{+}-J_{-}\phi_{-})]
\end{equation}
\end{widetext}
Taking the functional derivative we obtain the general amplitude
\begin{widetext}
\begin{equation}
\begin{array}{c}
\Bracket{\phi_{a_{1}}(\tau,x),...,\phi_{a_{N}}(\tau,x_{N})}=\\
\\
=\int\mathcal{D}\phi_{+}\mathcal{D}\phi_{-}\phi_{a_{1}}(\tau,x_{1})...\phi_{a_{N}}(\tau,x_{N}) \\
\\
\times exp[i\int_{\tau_{0}}^{\tau_{f}}d\tau d^{3}x(\mathcal{L}[\phi_{+}]-\mathcal{L}_{cl}[\phi_{-}]+J_{+}\phi_{+}-J_{-}\phi_{-})]=\\
\\
\frac{\delta}{i a_{1}\delta J_{a_{1}}(\tau,x_{1})}...\frac{\delta}{i a_{N}\delta J_{a_{N}}(\tau,x_{N})}
Z[J_{+},J_{-}]|_{J_{\pm}=0}
\end{array}
\end{equation}
\end{widetext}
Perturbative calculations are done in the standard fashion, splitting the Lagrangian into a free part and an interaction part. Then we can write the partition function as
\begin{widetext} 
\begin{equation}
Z[J_{+}, J_{-}]=exp[i\int_{\tau_{0}}^{\tau_{f}}d\tau d^{3}x(\mathcal{L}_{int}[\frac{\delta}{i\delta J_{+}}]
-\mathcal{L}_{int}[\frac{\delta}{i\delta J_{-}}])]
\end{equation}
\end{widetext}
and 
\begin{widetext}
\begin{equation}
Z_{0}[J_{+},J_{-}]=\int \mathcal{D}\phi_{+}\mathcal{D}\phi_{-} exp[i \int_{\tau_{0}}^{\tau_{f}}d\tau d^{3}x(\mathcal{L}_{0}[\phi_{+}]-\mathcal{L}_{0}[\phi_{-}]+J_{+}\phi_{+}-J_{-}\phi_{-})]
\end{equation}
\end{widetext}
the last equation being a Gaussian. The first equation can be expanded in a perturbative series and it will generate all diagrammatic rules associated to the Schwinger Keldysh formalism. The advantage is that we can now treat far from equilibrium systems in a consistent manner, while the downside is that we have to take care of the various time ordered and anti-time ordered field components. At the tree level we will have to derive the two point functions playing the role of propagators of the form 
\begin{widetext}
\begin{equation}
-i\Delta_{ab}(\tau_{1},x_{1};\tau_{2},x_{2})=\frac{\delta}{i a \delta J_{a}(\tau_{1},x_{1})}\frac{\delta}{i b \delta J_{b}(\tau_{2},x_{2})}Z_{0}[J_{+},J_{-}]|_{J_{\pm}=0}
\end{equation}
\end{widetext}
where $a,b=\pm$. Combinations of indices give us four different types of propagators. 
\begin{widetext}
\begin{equation}
\begin{array}{c}
-i\Delta_{++}(\tau_{1},x_{1}; \tau_{2},x_{2})=\frac{\delta}{i\delta J_{+}(\tau_{1},x_{1})} \frac{\delta}{i\delta J_{+}(\tau_{2},x_{2})}Z_{0}[J_{+},J_{-}]|_{J_{\pm}=0}=\\
\\
=\int\mathcal{D}\phi_{+}\mathcal{D}\phi_{-}\phi_{+}(\tau_{1},x_{1})\phi_{+}(\tau_{2},x_{2})e^{i\int d\tau d^{3}x (\mathcal{L}_{0}[\phi_{+}]-\mathcal{L}_{0}[\phi_{-}])}=\\
\\
=\sum_{\alpha}\Bracket{\Omega|\alpha}\Bracket{\alpha|T\{\phi(\tau_{1},x_{1})\phi(\tau_{2},x_{2})\}|\Omega}=\\
\\
=\bra{\Omega}T\{\phi(\tau_{1},x_{1})\phi(\tau_{2},x_{2})\}\ket{\Omega}\\
\end{array}
\end{equation}
\end{widetext}
Similarly the rest of the propagators will be 
\begin{widetext}
\begin{equation}
\begin{array}{c}
-i\Delta_{--}(\tau_{1},x_{1};\tau_{2},x_{2})=\frac{-\delta}{i\delta J_{-}(\tau_{1},x_{1})} \frac{-\delta}{i \delta J_{-}(\tau_{2},x_{2})}Z_{0}[J_{+}, J_{-}]|_{J_{\pm}=0}=\\
\\
=\int \mathcal{D}\phi_{+}\mathcal{D}\phi_{-}\phi_{-}(\tau_{1},x_{1})\phi_{-}(\tau_{2},x_{2})e^{i\int d\tau d^{3}x(\mathcal{L}_{0}[\phi_{+}]-\mathcal{L}_{0}[\phi_{-}])}=\\
\\
=\sum_{\alpha}\bra{\Omega}\bar{T}\{\phi(\tau_{1},x_{1})\phi(\tau_{2},x_{2})\}\ket{\alpha}\Bracket{\alpha|\Omega}=\\
\\
=\bra{\Omega}\bar{T}\{\phi(\tau_{1},x_{1})\phi(\tau_{2},x_{2})\}\ket{\Omega}\\
\end{array}
\end{equation}
\end{widetext}
as well as for the mixed terms
\begin{widetext}
\begin{equation}
\begin{array}{c}
-i\Delta_{+-}(\tau_{1},x_{1};\tau_{2},x_{2})=\frac{\delta}{i\delta J_{-}(\tau_{1},x_{1})} \frac{-\delta}{i \delta J_{-}(\tau_{2},x_{2})}Z_{0}[J_{+}, J_{-}]|_{J_{\pm}=0}=\\
\\
=\int \mathcal{D}\phi_{+}\mathcal{D}\phi_{-}\phi_{+}(\tau_{1},x_{1})\phi_{-}(\tau_{2},x_{2})e^{i\int d\tau d^{3}x(\mathcal{L}_{0}[\phi_{+}]-\mathcal{L}_{0}[\phi_{-}])}=\\
\\
=\sum_{\alpha}\bra{\Omega}\phi(\tau_{2},x_{2})\ket{\alpha}\Bracket{\alpha|\phi(\tau_{1},x_{1})|\Omega}=\\
\\
=\bra{\Omega}\phi(\tau_{2},x_{2})\phi(\tau_{1},x_{1})\ket{\Omega}
\\
\end{array}
\end{equation}
\end{widetext}
and finally 
\begin{widetext}
\begin{equation}
\begin{array}{c}
-i\Delta_{-+}(\tau_{1},x_{1};\tau_{2},x_{2})=\frac{-\delta}{i\delta J_{-}(\tau_{1},x_{1})} \frac{\delta}{i \delta J_{+}(\tau_{2},x_{2})}Z_{0}[J_{+}, J_{-}]|_{J_{\pm}=0}=\\
\\
=\int \mathcal{D}\phi_{+}\mathcal{D}\phi_{-}\phi_{-}(\tau_{1},x_{1})\phi_{+}(\tau_{2},x_{2})e^{i\int d\tau d^{3}x(\mathcal{L}_{0}[\phi_{+}]-\mathcal{L}_{0}[\phi_{-}])}=\\
\\
=\sum_{\alpha}\bra{\Omega}\phi(\tau_{2},x_{2})\ket{\alpha}\Bracket{\alpha|\phi(\tau_{1},x_{1})|\Omega}=\\
\\
=\bra{\Omega}\phi(\tau_{1},x_{1})\phi(\tau_{2},x_{2})\ket{\Omega}

\end{array}
\end{equation}
\end{widetext}
We go then to momentum space and express the fields $\phi$ in terms of mode functions $u(\tau,k)$ and creation and annihilation operators. With that we re-write the momentum propagators from the coordinate propagators above
\begin{equation}
G_{ab}(k, \tau_{1}, \tau_{2})=-i\int d^{3}x e^{-ik\cdot x}\Delta_{ab}(\tau_{1},x;\tau_{2},0)
\end{equation}
The tree level propagators in this formalism for the 3-momentum space become 
\begin{widetext}
\begin{equation}
\begin{array}{c}
G_{++}(k; \tau_{1},\tau_{2})=G_{>}(k; \tau_{1},\tau_{2})\theta(\tau_{1}-\tau_{2})+G_{<}(k,\tau_{1},\tau_{2})\theta(\tau_{2}-\tau_{2})\\
\\
G_{+-}(k;\tau_{1}, \tau_{2})=G_{<}(k,\tau_{1},\tau_{2})\\
\\
G_{-+}(k,\tau_{1},\tau_{2})=G_{>}(k,\tau_{2},\tau_{2})\\
\\
G_{--}(k,\tau_{1},\tau_{2})=G_{<}(k,\tau_{1},\tau_{2})\theta(\tau_{1}-\tau_{2})+G_{>}(k,\tau_{1},\tau_{2})\theta(\tau_{2}-\tau_{1})\\
\end{array}
\end{equation}
\end{widetext}
with 
\begin{equation}
\begin{array}{c}
G_{>}(k,\tau_{1},\tau_{2})=u(\tau_{1},k)u^{*}(\tau_{2},k)\\
\\
G_{<}(k,\tau_{1},\tau_{2})=u^{*}(\tau_{1},k)u(\tau_{2},k)\\
\end{array}
\end{equation}
there are only three of the four propagators that are linearly independent. Complex conjugation also gives us the relations $G^{*}_{>}=G_{<}$, $G_{--}=G^{*}_{++}$ and $G_{+-}^{*}=G_{-+}$. 
The associated Feynman diagrams will be constructed in a way that is very similar to the usual diagramatic expansion with the observation that now we have to take into account for a series of different propagators that do not appear in equilibrium approaches to quantum field theory. These will take into account the existence of field that are time ordered as well as those that are anti-time-ordered. Because of this the diagramatic rules will be more complex. The propagators will make the distinction between their two endpoints, involving time ordered or anti-time ordered fields. Also, we will make the distinction between the boundary points and bulk points. Here the boundary points appear in external legs as connecting to detectable states and the bulk points appear on both sides of internal legs that appear as inner propagators. A boundary point will not distinguish between a $+$ and a $-$ state and therefore there will only be two bulk to boundary propagators
\begin{equation}
\begin{array}{c}
G_{+}(k,\tau)=G_{++}(k,\tau,\tau_{f})\\
\\
G_{-}(k,\tau)=G_{-+}(k,\tau,\tau_{f})\\
\end{array}
\end{equation}
\section{4. The quantum Boltzmann equation and the Kadanoff Baym solution}
When describing open quantum systems and/or systems that are far from equilibrium we rely on the Schwinger Keldysh or on the Kadanoff Baym formalisms. The quantum Boltzmann equation is a quantum master equation for a statistical operator $\rho$ describing the motion of a distinguished test particle in a gas. In general it has the form 
\begin{equation}
\frac{d}{dt}\rho = \mathcal{M}\rho=\frac{1}{i\hbar}[H,\rho]+\mathcal{L}\rho
\end{equation}
where $H$ represents the hamiltonian of the test particle describing its kinetic energy as well as the variation in energy due to the presence of the background gas. As opposed to the classical Boltzmann equation, this formulation allows us to treat the quantum effects by standard quantum field theoretical approaches and to calculate the quantum density matrix directly. 
However, the usual Schwinger-Keldysh formalism is adapted for the case in which the initial density matrix is non-interacting and the evolution of the field operators is without interactions. This results in the absence of correlations in the initial phase. However, in order to properly describe the creation of entanglement in the first phase, we need to allow for a Kadanoff-Baym formalism [28], [29], [30] in which the initial context allows for correlations. This is exactly what we will do in the following chapters. 
Let us first start with a general initial time $t_{0}$ and consider that 
\begin{equation}
\rho(t_{0})=e^{-\beta H}/Z
\end{equation}
 where $H$ is the general interacting Hamiltonian. We consider now the equilibrium hamiltonian $H$ to be formed from a non-interacting part $H_{0}$ and the interactions encoded in a potential $V$. We wish to reformulate the interacting density matrix $e^{-\beta H}$ in terms of a non-interacting matrix $e^{-\beta H_{0}}$. We have the evolution operator 
 \begin{equation}
 U(t)=T exp(-i\int_{0}^{t} dt_{1}H(t_{1}))
 \end{equation}
 and we can write more generally the $S$-matrix as 
 \begin{equation}
 S(t,t')=e^{iHt}U(t,t')e^{-iHt'}=T exp(-i\int_{t'}^{t}dt_{1}H'(t_{1}))
 \end{equation}
 where we use imaginary time and in fact we introduce a general initial time $t_{0}$. The full equilibrium Hamiltonian is indeed $H$ and the interaction and Heisenberg pictures are defined with respect to our general initial time $t_{0}$
 \begin{equation}
A(t)=e^{i H(t-t_{0})}Ae^{-iH(t-t_{0})}
 \end{equation}
 and 
 \begin{equation}
 \hat{A}(t)=e^{i H_{0}(t-t_{0})}Ae^{-iH_{0}(t-t_{0})}
 \end{equation}
 The evolution operator is then $U(t,t_{0})=e^{-iH(t-t_{0})}$ and the S-matrix 
 \begin{equation}
 S(t,t_{0})=T exp(-i\int_{t'}^{t}dt_{1}V(t_{1}))
 \end{equation}
with $S(t,t_0)= e^{iH_{0}(t-t_{0})}U(t,t_{0})$. 
We have therefore 
\begin{equation}
e^{-\beta H}=e^{-\beta H_{0}}S(t_{0}-i\beta, t_{0})
\end{equation}
where the $S$ matrix $S(t_{0}-i\beta, t_{0})$ evolves the density matrix along a path on the imaginary axis $[t_{0}, t_{0}-i\beta]$. The Green function becomes 
\begin{widetext}
\begin{equation}
i G(x,t,x',t')=\frac{1}{Z} Tr e^{-\beta H_{0}}S(t_{0}-i\beta, t_{0})T_{c}[S_{c}(t_{0}, t_{0})\phi_{1}(x,t)\phi_{2}(x',t')]
\end{equation}
\end{widetext}
The field operators still evolve according to the full Hamiltonian. If we want to work in the interaction picture as we showed above, we have to transform them 
\begin{equation}
\hat{\phi}(x,t)=S(t_{0},t)\phi(x,t)S(t,t_{0})
\end{equation}
and we obtain 
\begin{widetext}
\begin{equation}
iG(x,t,x',t')=\frac{1}{Z}Tr e^{-\beta H_{0}}S(t_{0}-i\beta, t_{0}) T_{c}[S_{c}(t_{0},t_{0})S(t_{0},t)\hat{\phi}(x,t)S(t_{0},t')\hat{\phi}(t',x')S(t',t_{0})]
\end{equation}
\end{widetext}
where the hatted operators are those in the interaction picture. The $S$-matrix for interactions evolves along a three branch contour now $C^{*}=C_{0}\cup [t_{0},t_{0}-i\beta]$ which is the Kadanoff-Baym contour. This contour starts at $t_{0}$, evolves towards an arbitrary state through a path evolving towards time $max(t,t')$, returns returns back towards $t_{0}$ through a lower path at $min(t,t')$ but when reaching $t_{0}$ takes another downturn in the imaginary time towards $t_{0}-i\beta$. We define the ordering operator $T_{c^{*}}$ along this contour so that we can move the thermal S-matrix $S(t_{0}-i\beta, t_{0})$ into a contour ordered product. We then produce
\begin{widetext}
\begin{equation}
iG(x,t,x',t')=\frac{Tr e^{-\beta H_{0}}T_{c^{*}}[S_{c^{*}}(t_{0}-i\beta, t_{0})S_{c}(t_{0},t_{0})\phi(x,t)\phi(x',t')]}{Tr e^{-\beta H_{0}}T_{c^{*}}[S_{c^{*}}(t_{0}-i\beta,t_{0})S_{c}(t_{0},t_{0})]}
\end{equation}
\end{widetext}
with a contour ordered $S$-matrix defined along this Kadanoff-Baym contour

\begin{equation}
S_{c^{*}}(t_{0}-i\beta, t_{0})=T_{c^{*}}exp(-i\int_{C^{*}}d\tau_{1}\hat{V}(\tau_{1}))
\end{equation}

where the partition function is 
\begin{equation}
\begin{array}{c}
Z=Tr e^{-\beta H}=Tr e^{-\beta H_{0}}S(t_{0}-i\beta, t_{0})=\\
\\
=Tr e^{-\beta H_{0}}T_{c^{*}}[S_{c^{*}}(t_{0}-i\beta, t_{0})S_{c}(t_{0},t_{0})]\\
\\
\end{array}
\end{equation}
In this case we can apply Wick's theorem if we so wish and we can perform perturbative calculations, either by a path integral approach or by a second quantisation approach. 
This formalism must be used to be able to analyse initial correlations namely the effect of having an initial interaction at the level of the evolving particles on the later times $t>t_{0}$. 
The perturbation expansion for the contour ordered 1-particle Green's function on the non-equilibrium contour can be resummed in the form of an integral Dyson equation
\begin{widetext}
\begin{equation}
\begin{array}{c}
G(x,t,x',t')=G_{0}(x,t,x',t')+\int dx_{2}d\tau_{2}G_{0}(x,t,x_{1},\tau_{1})U(x_{2},\tau_{2})G(x_{2},\tau_{2},x',t')+\\
+\int dx_{2}d\tau_{2}\int dx_{3}d\tau_{3} G_{0}(x,t,x_{2},\tau_{2})\Sigma(x_{2},\tau_{2},x_{3},\tau_{3})G(x_{3},\tau_{3},x',t')
\end{array}
\end{equation}
\end{widetext}
where $G(x,t,x',t')=-i\Bracket{T_{c}[\phi(x,t)\phi(x',t')]}$ is the exact green function and $G_{0}(x,t,x',t')=-i\Bracket{T_{c}[\hat{\phi}(x,t)\hat{\phi}(x',t')]}$ is the unperturbed Green's function with field operators in the interaction picture, $U(x_{2},t_{2})$ is the 1-particle potential and $\Sigma(x_{2},\tau_{2},x_{3},\tau_{3})$ is the 1-particle irreducible self-energy. The above formula can be reduced to 
\begin{equation}
G=G_{0}+G_{0}U G+G_{0}\Sigma G
\end{equation}
The Keldysh theory sets $t_{0}\rightarrow -\infty$ and the initial correlations will be neglected. However, in the Kadanoff-Baym theory we have to include the initial correlations by keeping $t_{0}$ finite and assume that the initial density matrix $\rho(t_{0})$ has the equilibrium form 
\begin{equation}
\rho(t_{0})=e^{-\beta H}/Z
\end{equation}
where $H$ is an interacting Hamiltonian. Now, we can consider that $\rho(t_{0})$ is indeed not in equilibrium. This corresponds to a system prepared in a correlated non-equilibrium state in the beginning that will evolve over time without assuming that the evolution time will be much later than the initial time. For this, the original Kadanoff-Baym contour needs to be modified as well. The general non-equilibrium initial density matrix can be written as 
\begin{equation}
\frac{e^{-\lambda B(t_{0})}}{Tr e^{-\lambda B(t_{0})}}
\end{equation}
but in this case $\lambda$ will not play the role of a temperature and $B$ will not play the role of an Hamiltonian. We can however get the analogy with those quantities further and build a perturbation expansion of a modified Kadanoff-Baym contour with an imaginary region $[t_{0}-i\lambda, t_{0}]$. 
This approach will allow us to treat many-particle Green functions and initial correlations on the same footing being therefore able to construct the evolution of the density matrix with a given initial correlation. The Dyson equation will also be generalised accordingly. 
We can define an arbitrary operator in the Heisenberg picture as 
\begin{equation}
\mathcal{O}_{B}(\tau)=exp[i\tau B(t_{0})]\mathcal{O}_{S}(t_{0})exp[-i\tau B(t_{0})]
\end{equation}
Considering the density matrix $\rho$, we can take out a one-particle density matrix from it. We use for that a one-particle operator $B_{0}$ define d by 
\begin{equation}
B'=B-B_{0}
\end{equation}
where $B'$ is considered small. The density matrix can then be rewritten as 
\begin{equation}
\begin{array}{c}
\rho = \frac{1}{Z}exp(-\lambda B_{0}) exp(\lambda B_{0})exp(-\lambda B)=\\
=\frac{Z_{0}}{Z}\rho_{0} exp(\lambda B_{0})exp(-\lambda B)\\
\end{array}
\end{equation}
where 
\begin{equation}
\begin{array}{cc}
\rho_{0}=\frac{1}{Z_{0}}exp(-\lambda B_{0})\; & Z_{0}=Tr(exp(-\lambda B_{0}))
\end{array}
\end{equation}
is a one particle density matrix. The operator 
\begin{equation}
S_{c'}(\tau,0)=exp(i\tau B_{0})exp(-i\tau B)
\end{equation}
satisfies the differential equation
\begin{equation}
i\frac{\partial}{\partial \tau}S_{c'}(\tau,0)=B_{B_{0}}'(\tau)S_{c'}(\tau,0)
\end{equation}
Here $B_{B_{0}}'$ is given in the interaction picture with respect to $B_{0}$ which is defined for an arbitrary operator $O$ as 
\begin{equation}
O_{B_{0}}(\tau)=exp(i\tau B_{0})O_{S}(t_{0})exp(-i\tau B_{0})
\end{equation}
which after a formal integration leads to 
\begin{equation}
S_{c'}(\tau, 0)=T_{c'}exp[-i\int_{0}^{\tau}d\tau'B_{B_{0}}'(\tau)]
\end{equation}
where the time ordering operator provides us with an integration direction along the contour $c'=[0,-i\lambda]$. 
\begin{equation}
T_{c'}O_{1}(\tau_{1})O_{2}(\tau_{2})=O_{1}(\tau_{1})O_{2}(\tau_{2})
\end{equation}
for $\tau_{1}>\tau_{2}$ on $c'$
and
\begin{equation}
T_{c'}O_{1}(\tau_{1})O_{2}(\tau_{2})=-O_{2}(\tau_{2})O_{1}(\tau_{1})
\end{equation}
for $\tau_{1}\leq \tau_{2}$ on $c'$. 
With this and the explicit form of the density matrix factored out with the initial one
\begin{equation}
\rho=\frac{Z_{0}}{Z}\rho_{0}exp(\lambda B_{0})exp(-\lambda B)
\end{equation}
we obtain the initial density matrix 
\begin{equation}
\rho=\frac{Z_{0}}{Z}\rho_{0}S_{c'}(-i\lambda,0)
\end{equation}
We can now write expectation values at the initial time moment $t_{0}$ as 
\begin{equation}
\Bracket{O(t_{0})}=\frac{Z_{0}}{Z}\sum_{n_{0}}\Bracket{n_{0}(t_{0})|\rho_{0}S_{c'} O(t_{0})|n_{0}(t_{0})}
\end{equation}
We may regard $\rho_{0}$ as the initial density matrix and the product $\frac{Z_{0}}{Z}S_{c'}O(t_{0})$ as the operator to be averaged. This means we have expressed it in terms one particle density matrices and non-interacting operators hence and that makes the Wick theorem valid for each term in the expansion of the $S_{c'}$ matrix. For two point functions at times $t,t'\geq t_{0}$. We have to deal with the fact that the time development operator has to be a one particle operator. The evolution in time is described by a time-dependent Hamiltonian which usually includes many particle interactions. We need to expand the time-development operator similar to the treatment of the initial density matrix.
The two point function contains products of operators at two times. One of the components, an operator with a single time component in the Heisenberg picture will be written as
\begin{equation}
O_{\mathcal{H}}(t)=[\tilde{T}exp[i\int_{t_{0}}^{t}d\tau \mathcal{H}(\tau)]]O_{S}(t)[T\;exp[-i\int_{t_{0}}^{t}d\tau\mathcal{H}(\tau)]]
\end{equation}
where $T$ and $\tilde{T}$ are the time and anti-time ordering operators for real time. 
We can imagine two branches on the real line, one being the upper branch $t_{0}^{\wedge}=t_{0}$, the other one being the lower branch $t_{0}^{\vee}=t_{0}$. However both branches are the real line, just that one goes in one direction, while the other in the opposite. 
The expression given the contour ordering is
\begin{equation}
O_{\mathcal{H}}(t)=T_{c}exp[-i\int_{c}d\tau\mathcal{H}(\tau)]O_{S}(t)
\end{equation}
We decompose again the total hamiltonian into two parts
\begin{equation}
\mathcal{H}(t)=H_{0}+H'(t)
\end{equation}
where $H_{0}$ is a time independent non-interacting Hamiltonian. We can write operators in the interaction picture that obey the equation
\begin{equation}
i\frac{d}{dt}O_{H_{0}}(t)=[O_{H_{0}}(t),H_{0}]+i\frac{\partial}{\partial t}O_{H_{0}}(t)
\end{equation}
given the initial time $t_{0}$ as the time where the interaction and Heisenberg picture are equivalent because there the system has been first prepared, we have 
\begin{equation}
O_{\mathcal{H}}(t_{0})=O_{H_{0}}(t_{0})
\end{equation}
and the time evolution of the operators being 
\begin{equation}
O_{H_{0}}(t)=exp(i(t-t_{0})H_{0})O_{S}(t)exp(-i(t-t_{0})H_{0})
\end{equation}
resulting 
\begin{equation}
O_{\mathcal{H}}(t)=S_{c}(t_{0}^{\vee},t)O_{H_{0}}(t)S_{c}(t,t_{0}^{\wedge})
\end{equation}
where the S-matrix is now 
\begin{widetext}
\begin{equation}
S_{c}(\tau_{2},\tau_{1})=exp(i(\tau_{2}-t_{0})H_{0})\times T_{c}exp(-i\int_{\tau_{1}}^{\tau_{2}}d\tau \mathcal{H}(\tau))\times exp(-i(\tau_{1}-t_{0})H_{0})
\end{equation}
\end{widetext}
This S-matrix contains three evolution segments and it obeys the differential equation 
\begin{equation}
i\frac{\partial}{\partial \tau_{2}}S_{c}(\tau_{2},\tau_{1})=H^{'}_{H_{0}}(\tau_{2})S_{c}(\tau_{2},\tau_{1})
\end{equation}
which results to have the solution 
\begin{equation}
S_{c}(\tau_{2},\tau_{1})T_{c}exp(-i\int_{\tau_{1}}^{\tau_{2}}d\tau H^{'}_{H_{0}})
\end{equation}
We now generalise to an operator depending on two time arguments 
\begin{equation}
O_{\mathcal{H}}(t,t')=T_{c}S_{c}O_{H_{0}}(t,t')
\end{equation}
where the contour is chosen to go through $t$ and $t'$ in the ordering given by the operators $O_{1}(t)$ and $O_{2}(t')$. Now considering an initial density matrix we can write the expectation value of our two-time operator as
\begin{equation}
\Bracket{O(t,t')}=\frac{Z_{0}}{Z}\sum_{n_{0}}\Bracket{n_{0}(t_{0})|\rho_{0}S_{c'}T_{c}S_{c}O_{H_{0}}(t,t')|n_{0}(t_{0})}
\end{equation}
The initial density matrix has a twofold expansion and the time evolution operator is combined with it resulting in an expansion that is integrated over the total combined contour $c^{*}=c'c$ with an independent contour ordering operator $T_{c^{*}}$. Due to this expanded contour we also define a Heisenberg picture with respect to 
\begin{equation}
\mathcal{K}(\tau)=\mathcal{H}(\tau)
\end{equation}
for $\tau$ on $c$ and
\begin{equation}
\mathcal{K}(\tau)=\mathcal{B}(\tau)
\end{equation}
for $\tau$ on $c'$ 
which is a combination of Heisenberg pictures defined on the two contours. In the interaction picture we have 
\begin{equation}
K_{0}(\tau)=H_{0}(\tau)
\end{equation}
for $\tau$ on $c$ and
\begin{equation}
K_{0}(\tau)=B_{0}(\tau)
\end{equation}
for $\tau$ on $c'$.
and then we get the overall S-matrix
\begin{widetext}
\begin{equation}
S_{c^{*}}=S_{c^{*}}(-i\lambda, t_{0}^{\wedge})=S_{c'}(-i \lambda,0)S_{c}(t_{0}^{\vee}, t_{0}^{\wedge})=T_{c^{*}}exp(-i\int_{c^{*}}d\tau K'_{K_{0}}(\tau))
\end{equation}
\end{widetext}
where 
\begin{equation}
K'_{K_{0}}(\tau)=H'_{H_{0}}(\tau)
\end{equation}
with $\tau$ on $c$ and
\begin{equation}
K'_{K_{0}}(\tau)=B'_{B_{0}}(\tau)
\end{equation}
with $\tau$ on $c'$
and then we obtain the expectation value for our two-times operator as 
\begin{equation}
\Bracket{O(t,t')}=\frac{Z_{0}}{Z}_{0}\Bracket{T_{c^{*}}S_{c^{*}}O_{H_{0}}(t,t')}_{0}
\end{equation}
where we used the notation 
\begin{equation}
_{0}\Bracket{O}_{0}=Tr(\rho_{0}O)
\end{equation}
Therefore we arrived at a representation where we use a one-particle density matrix $\rho_{0}$ and all operators are given in the interaction picture corresponding to a non-interacting system. This means that the Wick theorem holds for each term in the expansion of the $S$ matrix independently. However, we will also obtain in this expansion functions that are not real-time correlations. The system of equations is therefore not closed with respect to the real-time functions. We can obviously understand this by thinking that the operator $O_{H_{0}}(t,t')$ is defined on the real time contour while the $S_{c^{*}}$ matrix is defined on the full contour $c^{*}$. During the contraction we find averages over all possible pairings of two operators involving all possible positions on the contour $c^{*}$ and therefore we have to extend $O_{H_{0}}$ to $c^{*}$. We do that by constructing the $c^{*}$-contour ordered Green's function

The final expression will contain a $3\times3$ matrix representation that includes the Keldysh formalism as an upper left corner of the matrix. If the statistical average is taken over the one particle density matrix then $B'$ vanishes since no expansion of $\rho$ is necessary and the vertices $V_{B_{0}}$ which connect the Keldysh matrix to the correlators $\hat{G}^{13}$, $\hat{G}^{23}$, $\hat{G}^{31}$, $\hat{G}^{32}$ are zero and the Dyson equation for the $3\times 3$ matrix decouples into a Dyson equation for the usual $(2\times 2)$ Keldysh matrix and a Dyson equation for $\hat{G}^{33}$. 

\section{5. The Quantum Boltzmann Equation}
One instrument to be used in this article will be the quantum Boltzmann equation. Introductory discussions about it can be found in [25], [26], [27]. Its classical counterpart describes the statistical behaviour of a thermodynamic system that is not in a state of equilibrium nor does it have to be near a state of equilibrium. The equation appears in the analysis of the probability distribution of the positions and momenta of a typical, single out particle, namely the probability that a particle is occupying a single, determined, very small volume of a 6 dimensional space (a region around a position $r$ and a momentum $p$). The result is a non-linear integro-differential where the unknown is the probability density function in the six dimensional space of particle position and momentum. If we call the probability per unit phase space volume for $N$ particles to be in the phase space volume, to be $f(r,p,t)$ at an instant $t$ then 
\begin{equation}
dN=f(r,p,t)d^{3}r d^{3}p
\end{equation}
is the number of particles which all have positions lying within a volume element of the phase space. Integrating over a region in phase space we obtain the total number of particles with position and momenta in that region 
\begin{equation}
N=\int d^{3}p\int d^{3}r f(r,p,t)
\end{equation}
However, it is important to notice that while the probability density function $f(r,p,t)$ is associated to a number of particles, the phase space description is for one-particle states only, and doesn't directly consider the many particle relations. The change in the probability density function will therefore be described by three terms
\begin{equation}
\frac{df}{dt}=(\frac{\partial f}{\partial t})_{force} + (\frac{\partial f}{\partial t})_{diff}+(\frac{\partial f}{\partial t})_{coll}
\end{equation}
The first term represents the external influence on the particles, for example an external magnetic field, the 
"diff" term represents the diffusion term for the particles, and the last term is the term produced by the collision between particles. The total differential of $f$ is 
\begin{equation}
df=\frac{\partial}{\partial t}dt + \nabla f\cdot \frac{p}{m}dt+\frac{\partial f}{\partial p}\cdot F dt
\end{equation}
where $F$ would encode the external force and $p$ is the particle momentum. If we take also collisions into account the end result will be 
\begin{equation}
\frac{\partial f}{\partial t}+\frac{p}{m}\cdot \nabla f+F\cdot \frac{\partial f}{\partial p}=(\frac{\partial f}{\partial t})_{coll}
\end{equation}
Clearly, translating such an equation in the quantum language is not an easy task. An adaptation of this equation describes the motion of a single distinguished test particle affected by collisions with an ideal stationary background gas. What we generally describe however is the evolution of a probability density function or probability distribution defined on the phase space of a single particle. For the Boltzmann equation this is the marginal one-particle distribution of a dilute gas whose self-interaction gives rise to the non-linear form of the integro-differential equation. If we consider only one single test particle instead of a volume element, no such non-linearity exists as we exclude the self-interaction. However, in a relativistic context such linearisation cannot truly be considered due to quantum field theoretical effects. In any case, the equation is derived by means of a Stosszahlansatz in a similar way. The coupling of the environmental gas with the distinguished particle can be strong and the distinguished particle may be very far fro equilibrium. The background gas however is taken to be in a stationary state, usually at thermal equilibrium. Such equations describing open systems (systems in interaction with some background) that may be far from equilibrium are called master equations. Usually we employ Boltzmann equations when the kinetic description is based on interaction events that can be modelled as scattering processes. It seems therefore natural that the first quantum extension of this formula involved quantum corrections of the quantum statistics on the classical collision terms. Both the classical and quantum formulations include the two body interaction between test particle and gas particle in a fundamentally non-perturbative fashion by means of scattering theory. The classical theory contains the differential cross section to incorporate such collisions, the quantum master equation relies on pairs of scattering amplitudes. These objects have a complex phase and appear as operator valued quantities if the master equation is written in a representation independent form. The quantum Boltzmann equation is capable of describing the relation between two distinct gas induced phenomena occurring on different time scales and usually requiring some non-perturbative description of the coupling to the gas. First we have the collisional decoherence effect, which occurs at short times when the energy exchange between the gas and the particle can be ignored. Second, the frictional effect experienced by the test particle at longer time scales leading to a slowing down of the particle and eventually to thermalisation. For example, if a particle starts in a delocalised quantum state, the equation should be able to predict how it gradually loses this quantum property due to the dissemination of positional information by means of gas collisions. In the limit, as the quantum properties of the particle become less and less obvious one should obtain a relaxation and equilibration behaviour similar to the one predicted by the classical Boltzmann equation. Therefore a Boltzmann type description is fundamental in understanding the evolution of entanglement as a resource in the context of photons interacting with each other whether this interaction is described by a square diagram of electron-positron creation or by the presence of interaction mediating gravitons. Moreover, such a description may be useful in describing experiments in which high intensity photon beams are used to create entanglement between given many photon states on two sides. The fact that gravitons can mediate this type of entanglement can therefore be directly studied. Recently it has become possible to study the influence of a gas environment on the motion of a distinguished particle which is very far from a classical state, for example as a superposition of macroscopically distinct spatial positions. This also allowed to study quantum decoherence effects independently from dissipative effects. We can write the quantum Boltzmann equation for a density operator as 
\begin{equation}
\begin{array}{c}
\frac{d}{dt}\rho=\mathcal{M}\rho=\\
=\frac{1}{i\hbar}[H_{0}+H_{n}(P),\rho]+\mathcal{L}\rho\\
\end{array}
\end{equation}
where $H{0}=\frac{P^{2}}{2M}$ is the kinetic energy of the test particle and $H_{n}(P)$ is the energy shift due to the presence of the background gas. The superoperator $\mathcal{L}$ is a linear map which considers the incoherent part of the collisions with the gas. There have been several ways to express it, one of them being the Lindblad form as 
\begin{widetext}
\begin{equation}
\mathcal{L}\rho=\int dQ \int_{Q^{\perp}}dk_{\perp}[e^{iQ\cdot X/\hbar}L(k_{\perp},P;Q)\rho L^{\dagger}(k_{\perp},P;Q)e^{-iQ\cdot X/\hbar}-\frac{1}{2}\{\rho, L^{\dagger}(k_{\perp},P;Q)L(k_{\perp},P;Q)\}]
\end{equation}
\end{widetext}
where $X$ and $P$ are the position and momentum of the test particle and the anti-commutator is described by $\{-\}$. The integration is performed over momentum transfers $Q$ and over gas particle momenta from the plane perpendicular to $Q$. The function $L$ is operator valued and contains all the collisional interaction details with the gas. It has the expression 
\begin{widetext}
\begin{equation}
L(p,P;Q)=\sqrt{\frac{n_{gas}m}{m_{*}^{2}Q}}f(rel(p_{\perp Q},P_{\perp Q})-\frac{Q}{2}, rel(p_{\perp Q}, P_{\perp Q})+\frac{Q}{2}) \sqrt{\mu(p_{\perp Q}+\frac{m}{m_{*}}\frac{Q}{2}+\frac{m}{M}P_{\parallel Q})}
\end{equation}
\end{widetext}
Here we have $m$ and $M$ the masses of the gas particles and the test particle, respectively, and $m_{*}=mM/(m+M)$ is the reduced mass. The gas number density is $n_{gas}$ and the distribution function of the gas momenta is $\mu(p)$. The scattering amplitude determined by the two body interaction between the gas and the test particle is $f(p_{f},p_{i})$. The subscripts $\parallel$ and $\perp$ represent the component of the vector parallel and perpendicular to the momentum transfer $Q$. The relative momenta are defined by 
\begin{equation}
rel(p,P)=\frac{m_{*}}{m}p-\frac{m_{*}}{M}P
\end{equation}
The Hamiltonian correction is 
\begin{equation}
H_{n}(P)=-2\pi \hbar^{2}\frac{n_{gas}}{m_{*}}\int dp \mu(p) Re[f(rel(p,P), rel(p,P))]
\end{equation}
which is the energy shift due to forward scattering. To establish the relation to the classical correspondent of the quantum Boltzmann equation we can express it in the form of improper momentum Eigenstates 
\begin{equation}
\hat{P}\ket{P}=P\ket{P}
\end{equation}
the expression of the collision term will then be
\begin{widetext}
\begin{equation}
\bra{P}\mathcal{L}\ket{P'}=\int dQM_{in}(P,P';Q)\bra{P-Q}\rho\ket{P'-Q}-\frac{1}{2}[M_{out}^{cl}(P)+M_{out}^{cl}(P')]\bra{P}\rho\ket{P'}
\end{equation}
\end{widetext}
The gain term is given by 
\begin{equation}
M_{in}(P,P';Q)=\int_{Q^{\perp}}dk_{\perp} L(k_{\perp}, P-Q; Q)L^{*}(k_{\perp},P'-Q;Q)
\end{equation}
This reduces to the rate known from the classical Boltzmann equation if one approaches the diagonal $P=P'$. Using the quantum mechanical differential cross section $\sigma(p_{f},p_{i})=|f(p_{f},p_{i})|^{2}$ is given by 
\begin{widetext}
\begin{equation}
M_{in}^{cl}(P+Q;Q)=\frac{n_{gas}m}{m_{*}^{2}Q}\int_Q^{\perp}dk_{\perp}\mu(k_{\perp}+\frac{m}{m_{*}}\frac{Q}{2}+\frac{m}{M}P_{\parallel Q}) \sigma(rel(k_{\perp},P_{\perp Q})-\frac{Q}{2}, rel(k_{\perp},P_{\perp Q})+\frac{Q}{2})
\end{equation}
\end{widetext}
which means it gives the rate of collisions that changes the test particle momentum from $P_{i}$ to $P_{f}$. 
The loss term is given by the arithmetic mean of the loss rate which appears in the corresponding classical equation. The latter is related to the classical gain rate 
\begin{equation}
M_{out}^{cl}=\int dQ M_{in}^{cl}(p+Q;Q)
\end{equation}
and therefore the probability will be conserved in the classical case. 
If we only consider the diagonal term we immediately recover the classical Boltzmann equation if one replaces the scattering amplitudes by their Born approximations and results in the removal of the operator-valuedness of the scattering amplitudes. 
The quantum Boltzmann equation emerges from the condition of compatibility with the classical equation in the suitable limit and from the demand of having the general form of a covariant, completely positive master equation. The classical Boltzmann equation describes the collision dynamics of a test particle interacting with a homogeneous background gas described by a momentum distribution $\mu(p)$. The non-trivial part of the equation is therefore the one that contains the collision contribution to the time evolution of the distribution function $f(P)$, which we denote by $\partial_{t}^{coll} f(P)$
\begin{widetext}
\begin{equation}
\partial_{t}^{coll}f(P)=\frac{n_{gas}}{m_{*}^{2}}\int dP' \int dp' \int dp \sigma(rel(p,P),rel(p',P'))\delta(\frac{P'^{2}}{2M}+\frac{p'^{2}}{2m}-\frac{P^{2}}{2M}-\frac{p^{2}}{2m})\delta^{3}(P'+p'-P-p)[\mu(p')f(P')-\mu(p)f(P)]
\end{equation}
\end{widetext}
where the delta functions encode the conservation of energy and momentum in each single collision. Evaluating the two delta functions we obtain 
\begin{widetext}
\begin{equation}
\begin{array}{c}
\partial_{t}^{coll}f(P)=\frac{m_{gas}m}{m_{*}^{2}}\int\frac{dQ}{Q}\int_{Q^{\perp}}dk_{\perp}\sigma(rel(k_{\perp},P_{\perp Q})-\frac{Q}{2}, rel(k_{\perp}, P_{\perp Q}+\frac{Q}{2})\times\\
\times [\mu(k_{\perp}+\frac{m}{m_{*}}\frac{Q}{2}+\frac{m}{M}(P_{\parallel Q}-Q))f(P-Q)-\mu(k_{\perp}+\frac{m}{m_{*}}\frac{Q}{2}+\frac{m}{M}P_{\parallel Q})f(P)]
\end{array}
\end{equation}
\end{widetext}
or, using the previous notation 
\begin{equation}
\partial_{t}^{coll}f(P)=\int dQ M_{in}^{cl}(P;Q)f(P-Q)-M_{out}^{cl}(P)f(P)
\end{equation}
and hence we identified the classical momentum distribution with our density matrix operator 
$f(P)\cong \bra{P}\rho\ket{P}$. The requirement that we obtain back the classical equation in the proper limit is generally satisfied, however this requirement is not sufficient to fix the quantum master equation which is required to describe the coherences corresponding to off-diagonal elements. 
The completion of an expression of the Boltzmann equation valid for the quantum counterpart is no easy task. If one assumes a Markovian structure for the master equation then a Lindblad form must be used to fix it
\begin{equation}
\mathcal{L}\rho = \sum_{j}[L_{j}\rho L_{j}^{\dagger}-\frac{1}{2}\{L_{j}^{\dagger}L_{j},\rho\}]
\end{equation}
However, it is well known that such a Markovian approach doesn't fully describe quantum dynamics. 
A more modern approach, based on the translational covariance of $\mathcal{L}\rho$ has been provided in ref. [16]. Early work on this subject has led to the introduction of quantum dynamical semigroups [17]. A final characterisation was obtained by Holevo [18,19,20,21] constructing on a quantum generalisation of the classical Levy-Khintchine formula. If the background gas is considered to be homogeneous the resulting Boltzmann equation, both classical and quantum, is considered to be invariant under translations and the interactions between colliding particles are described by a two-body potential. In the quantum version this leads to the master equation being translation covariant and hence its action must commute with the unitary representation of translations
\begin{equation}
\mathcal{L}(e^{-iA\cdot {/\hbar}}\rho e^{i A\cdot P/\hbar})=e^{-iA\cdot P/ \hbar} \mathcal{L}\rho e^{iA\cdot P/ \hbar}
\end{equation}
The structure of a master equation which obeys these rules is given by 
\begin{widetext}
\begin{equation}
\mathcal{L}\rho=\int dQ\sum_{j}[e^{i Q\cdot X/ \hbar}L_{j}(P;Q)\rho L_{j}^{\dagger}(P;Q)e^{-iQ\cdot X/\hbar}-\frac{1}{2}\{L_{j}^{\dagger}(P;Q)L_{j}(P;Q),\rho\}]
\end{equation}
\end{widetext}
There still remains freedom in the determination of the quantum Boltzmann equation, and this arbitrariness requires the fixing of the non-diagonal terms. The mapping $\mathcal{L}$ has not been fixed precisely for the off-diagonal terms. 

The symmetry properties of the quantum Boltzmann equations to translation operations, rotations, or relativistic boosts are shared with the classical Boltzmann equations. So are also the relaxation properties and the existence of a stationary solution. Those are all characteristics of the diagonal matrix elements. The interference phenomena which are relevant when the motion of the test particle may be characterised by a coherently delocalised state are described by the off-diagonal terms and are important for the following description. For example the dynamical transition of an initially delocalised superposition state into an incoherent particle-like mixture can be described also by means of the quantum Boltzmann equation but its information is given by the non-diagonal matrix elements. For that description the coherent part and its loss term are most important. The incoherent part was described by the $\mathcal{L}$ operator while the modified hamiltonian $H=\frac{P^{2}}{2M}+H_{n}(P)$ describes the coherent part. The supplemental term $H_{n}$ describes the modification of the quantum wavefunction dispersion relation due to coherent interactions with the background particles. This energy shift is related to the forward scattering process which does not change the momentum of the background particles and of the test particle. The forward scattering amplitude $f_{0}(p)=f(p,p)$ is calculated as a thermal average over the momentum distribution function of the background system
\begin{equation}
\Bracket{f_{0}(P)}_{\mu}=\int dp \mu(p) f(rel(p,P),rel(p,P))
\end{equation}
leading to 
\begin{equation}
H_{n}(P)=-2\pi h^{2}\frac{n_{gas}}{m_{*}}Re\Bracket{f_{0}(P)}_{\mu}
\end{equation}
This type of forward scattering may be observed in detectors as a phase shift in the interference pattern. If particles do not get lost (open system) then their motional state is modified at most coherently in that situation. The effect of particle loss is included by the atenuation of the amplitude. 
In general, the combined state of the system and the environment is no longer separable after the interaction, resulting in a loss of purity in the reduced quantum state, after tracing the environmental state. The interaction transfers the information from the system to the environment. 

However, in the early stage of the interaction, quantum correlation is rapidly created, due to the inner mechanisms of the interaction, that are not accounted for strictly in the construction of the Boltzmann equation. Therefore, to properly describe the evolution of entanglement, one has to combine the quantum Boltzmann equation with the proper description of the graviton mediated photon-photon interaction, as has been shown in the previous chapter. One therefore expects that if the test particle is initially in a superposition of two positions as in an interferometer, the coherence of this state will be reduced more, if the scattering event is able to better distinguish between the two positions leading to the extraction of the positional information so that it becomes available to an external observer. 
However, superposition itself is not entanglement. In the case of entanglement, the information about the position of one particle is shared globally (inseparably) with another particle (say a particle in the gas). 

Finally, to properly describe the quantum effects by means of proper quantum field theoretical approaches, we need to consider the descriptions far from equilibrium and their associated amplitudes, but this has already been done in the previous chapter where I described the Kadanoff-Baym formulation. Using those results together with the unrestricted form of the quantum Boltzmann equation we obtain a more general quantum description in which entanglement can be dynamically followed. With that we can now move forward and obtain the evolution of photons, interacting via gravitons, in a photon gas. 
Such a description is particularly suitable for the construction of early cosmological phenomenology relating the interaction between photons mediated by gravitons, resulting in entangled pairs, but also in the study of the evolution of entanglement in particles emerging from pair creation in high intensity laser beams. Having high intensity laser beams in mind, we may also consider analysing the entanglement dynamics for photons in high intensity photon beams.
Using the Lindblad operators as above, expanding the exponentials associated to the invariance of the equation to translation transformations and calculating the calculating the averages in the second quantisation we arrive at the following form of the quantum Boltzmann equation
\begin{widetext}
\begin{equation}
(2\pi)^{3}\delta^{3}(0)(2k^{0})\frac{d\rho_{ij}^{\gamma}(k)}{dt}=i\Bracket{[H_{\gamma g}(t), \mathcal{D}_{ij}^{\gamma}(k)]}-\frac{1}{2}\int_{-\infty}^{\infty}dt'\Bracket{[H_{\gamma g}(t'),[H_{\gamma g}(t'),[H_{\gamma g}(t),\mathcal{D}_{ij}^{\gamma}(k)]]]}
\end{equation}
\end{widetext}
where off diagonal terms are properly considered in the expansion and the operator $D_{ij}$ is associated to the number of particles operator appearing from the interaction amplitude construction.
Next we explain the probability amplitudes from a QFT point of view, using the far from equilibrium description of the previous chapter. 

\section{6. Photons scattering on gravitons}
We will describe the scattering of photons on gravitons by the exchange of a graviton.
Starting with the Einstein Hilbert action 
\begin{equation}
S_{EH}=\frac{2}{k^{2}}\int d^{4}x\sqrt{-g}\mathcal{R}
\end{equation}
where $\mathcal{R}=R^{\lambda}_{\;\;\mu\lambda\nu}g^{\mu\nu}$ is the Ricci scalar curvature, 
we consider the weak field approximation and in particular the development of the gravitational perturbation around the Minkowski vacuum, we can write 
\begin{equation}
g_{\mu\nu}=\eta_{\mu\nu}+k\cdot h_{\mu\nu}
\end{equation}
Expanding the determinant of the metric according to this formula we obtain 
\begin{equation}
\sqrt{-g}=\sqrt{-\det(g)}=exp(\frac{1}{2}Tr(log(g)))=1+\frac{1}{2}\eta^{\mu\nu}h_{\mu\nu}+...
\end{equation}
To describe the matter sector of the problem (including our oscillators and the photons) we use the energy momentum tensor, which can be defined starting with the lagrangian as
\begin{equation}
T_{\mu\nu}=\frac{2}{\sqrt{-g}}\frac{\delta\sqrt{-g}\mathcal{L}_{mat}}{\delta g^{\mu\nu}}
\end{equation}
and then the interaction or coupling term between matter and gravity will be 
\begin{equation}
\mathcal{L}_{int}=-\frac{k}{2}h_{\mu\nu}T^{\mu\nu}
\end{equation}
If one starts with the Maxwell Lagrangian density $\mathcal{L}_{EM}=-\frac{1}{4}F_{\alpha\beta}F^{\alpha\beta}$ then using that $\delta\sqrt{-g}=-\frac{1}{2}g_{\alpha\beta}\delta g^{\alpha\beta}$ and we obtain 
\begin{equation}
\begin{array}{c}
T_{\alpha\beta}=\frac{1}{2}(F_{\mu\nu}F_{\rho\sigma}\frac{\delta(g^{\mu\nu}g^{\nu\sigma})}{\delta g^{\alpha\beta}}-\frac{1}{2}g_{\alpha\beta}F_{\delta\gamma}F^{\delta\gamma})=\\
\\
=F_{\alpha\lambda}F^{\lambda}_{\beta}-\frac{1}{4}g_{\alpha\beta}F_{\delta\gamma}F^{\delta\gamma}
\end{array}
\end{equation}
Such a stress tensor is conformally invariant, $T_{\mu}^{\;\;\mu}=0$. 
The gravitational fluctuations can be promoted to quantum operators following a standard quantum field theoretical approach 
\begin{equation}
\hat{h}_{\mu\nu}=\mathcal{A}\int d^{3}k\sqrt{\frac{\hbar}{2\omega_{k}(2\pi)^{3}}}(\hat{P}_{\mu\nu}(k)e^{-i k\cdot r}+H.c.)
\end{equation}
The operators $\hat{P}_{\mu\nu}$ and $\hat{P}^{\dagger}_{\mu\nu}$ are the gravitational annihilation and creation operators. The interaction term of the matter field to the graviton is represented by the operator valued distribution
\begin{equation}
\hat{H}_{int}=-\frac{1}{2}\int dr \hat{h}^{\mu\nu}\hat{T}_{\mu\nu}(r)
\end{equation}
Looking at the graviton, in the weak coupling regime around the Minkowski vacuum we can write 
\begin{equation}
\hat{h}_{\mu\nu}=\hat{\gamma}_{\mu\nu}-\frac{1}{2}\eta_{\mu\nu}\hat{\gamma}
\end{equation}
where $\gamma=\eta_{\mu\nu}\cdot \gamma^{\mu\nu}$.
We obtain two different modes from this expansion, one that behaves like a spin-0 object, the other like a spin-2 object. By the standard quantisation prescription we can promote them to operators and we obtain 
\begin{equation}
\begin{array}{c}
\hat{\gamma}_{\mu\nu}=A\int d^{3}k\sqrt{\frac{\hbar}{2\omega_{k}(2\pi)^{3}}}(\hat{P}^{\dagger}_{\mu\nu}(k)e^{-ik\cdot r}+H.c.)\\
\\
\hat{\gamma}=A\int d^{3}k\sqrt{\frac{\hbar}{2\omega_{k}(2\pi)^{3}}}(\hat{P}^{\dagger}(k)e^{-ik\cdot r}+H.c.)\\
\end{array}
\end{equation}
leading to the Hamiltonian 
\begin{equation}
\hat{H}_{g}=\int d^{3}k\hbar\omega_{k}(\frac{1}{2}\hat{P}_{\mu\nu}^{\dagger}(k)\hat{P}^{\mu\nu}(k)-\hat{P}^{\dagger}(k)\hat{P}(k))
\end{equation}
On the matter side we have the matrix elements of the stress energy tensor of photons. These are symmetric and correspond to the conserved energy momentum tensor $T_{\mu\nu}$ in flat spacetime

\begin{equation}
\begin{array}{c}
\Bracket{0|T_{\mu\nu}(x)|k'k}=e^{-i(k+k')\cdot x} \Bracket{0|T_{\mu\nu}(0)|k'k},\\
\\
T_{\mu\nu}=T_{\nu\mu},\\
\\
\partial_{\mu}T^{\mu\nu}=0\\
\end{array}
\end{equation}
The described phenomenon is that of two incoming photons (zero mass, spin 1) both taken on shell, therefore having 
\begin{equation}
\begin{array}{cc}
k^{2}=k'^{2}=0, & \epsilon\cdot k = \epsilon'\cdot k = \epsilon'\cdot k' =0
\end{array}
\end{equation}
where  $\epsilon$ and $\epsilon'$ are polarisations associated up to a gauge choice to the two photons of momenta $k$ and $k'$. 
Taking the Fourier transform and requiring momentum conservation we obtain 
\begin{equation}
\begin{array}{c}
\Bracket{0|T_{\mu\nu}(0)|k'k}N=\Bracket{0|T_{\mu\nu}(0)|k'k}_{tree} N F_{1}(t)+\\
\\
P_{\mu\nu}(q)[2(\epsilon'\cdot k)(\epsilon\cdot k')-q^{2}(\epsilon\cdot \epsilon')]F_{2}(t)\\
\\
+p_{\mu}p_{\nu}[2(\epsilon'\cdot k)(\epsilon\cdot k')-q^{2}(\epsilon\cdot \epsilon')]F_{3}(t)\\
\\
\\
\end{array}
\end{equation}
with $p=k'-k$, $q=-(k+k')$, $t=q^{2}=2k\cdot k'$, $P_{\mu\nu}(q)=q_{\mu}q_{\nu}-\eta_{\mu\nu}\cdot q^{2}$
and the normalisation factor $N$.
The first term has the form 
\begin{widetext}
\begin{equation}
\Bracket{0|T_{\mu\nu}(0)|k'k}_{tree} N = (k_{[\mu}\epsilon_{\alpha]}k'_{[\nu}\epsilon'_{\beta]}+k_{[\nu}\epsilon_{\alpha]}k'_{[\mu}\epsilon'_{\beta]})\eta^{\alpha\beta}-\frac{1}{2}\eta_{\mu\nu}k^{[\alpha}\epsilon^{\beta]}k'_{[\alpha}\epsilon'_{\beta]}
\end{equation}
\end{widetext}
The first term above corresponds to the free photon, being a classical term, then we have the identically conserved terms $P_{\mu\nu}$ associated to the improvement terms, and then the projector terms $p_{\mu}p_{\nu}$, orthogonal to $q_{\mu}$ and conserved by the on-shell condition. 
The matrix element obtained in this way depends on helicity as shown in the following matrix
\begin{widetext}
\begin{equation}
\Bracket{0|T^{\mu\nu}(0)|k'^{h'}k^{h}}N=
\begin{bmatrix}
    \frac{1}{2}\Bracket{k'\sigma^{\mu}k}\Bracket{k'\sigma^{\nu}k}F_{1}(t)       & -\Bracket{k k'}^{2}(P_{\mu\nu}(q)F_{2}(t)+p^{\mu}p^{\nu}F_{3}(t)) \\
    -[kk']^{2}(P^{\mu\nu}(q)F_{2}(t)+p^{\mu}p^{\nu}F_{3}(t))       & \frac{1}{2}\Bracket{k\sigma^{\mu}k'}\Bracket{k\sigma^{\nu}k'}F_{1}(t)  \\
\end{bmatrix}
\end{equation}
\end{widetext}
The rows and columns of the matrix are organised according to helicity and hence the diagonal terms are helicity preserving, basically preserving the opposite helicity of the two photon states $h=-h'$, while the off-diagonal terms are flipping the helicity leading to states of the form $h=h'$. We can see that the helicity preserving terms are governed by $F_{1}(t)$ and hence by the classical part while the helicity flipping terms are governed by $F_{2}(t)$ and $F_{3}(t)$. Therefore in a scattering process against an off-shell graviton we would obtain a dominant classical helicity preserving term, as well as an overlap between the helicity flipping photon pair with zero spin in the direction of motion and either a spin-0 or a spin-2 state described by $T_{\mu\nu}\ket{0}$. 
The gravitational helicity flipping is a well known phenomenon regarding the transfer of angular momentum from a large gravitational body to photons scattering off of it. However, the same effect appears in the process of mediating the interaction between two bodies by means of gravitons. In particular, such would be a non-classical effect to be recovered before taking into account loop corrections to the stress energy tensor. That would amount to a new way of measuring how exactly the gravitons could couple to photons and entangle them. 
In general we could construct an interaction term of the form 
\begin{equation}
\hat{H}_{int}=\frac{1}{2}\int dr [\hat{\gamma}_{\mu\nu}+\frac{1}{2}\hat{\gamma}(r)]\hat{T}_{\mu\nu}(r)
\end{equation}
The transfer of helicity from one photon to the other via the graviton exchange cannot be explained strictly classically. Also, it means that gravity can couple the helicity of the two photons leading to an entangled state. This effect also cannot be explained assuming a classical form for the graviton. 
Let us analyse the helicity states of the photon in more detail. The time evolution of the polarisation matrix of a photon is described by the quantum Boltzmann equation.
Using the quantum Boltzmann equation for the analysis of this problem allows us to study the evolution of entanglement in time and to take into account the system of photons as an open system describing the photon gas in the early universe. Indeed, the application of this prescription is being used in cosmology but due to the difficulties of obtaining any relevant quantum effects, this type of reasoning was impossible before the development of quantum Boltzmann equations where the function to be determined is the quantum density matrix [10-15]. 
The lowest term in the perturbative expansion of the power series of the interaction Hamiltonian of the theory will be the so called forward scattering term and will provide the most important effects in the dynamics of the photon polarisation. 
The Boltzmann equation for a photon interacting with a graviton is 
\begin{widetext}
\begin{equation}
(2\pi)^{3}\delta^{3}(0)(2k^{0})\frac{d\rho_{ij}^{\gamma}(k)}{dt}=i\Bracket{[H_{\gamma g}(t), \mathcal{D}_{ij}^{\gamma}(k)]}-\frac{1}{2}\int_{-\infty}^{\infty}dt'\Bracket{[H_{\gamma g}(t'),[H_{\gamma g}(t'),[H_{\gamma g}(t),\mathcal{D}_{ij}^{\gamma}(k)]]]}
\end{equation}
\end{widetext}
we have used the notation $\rho_{ij}^{\gamma}$ for the photon polarisation, $\mathcal{D}_{ij}^{\gamma}$ for the photon number operator, and $H_{\gamma g}(t)$ as the interaction Hamiltonian between photons and gravitons. On the right hand side of the above equation the first term is the so called forward scattering term, while the second one is the damping term. Keeping only the forward scattering term results in the following Boltzmann equation
\begin{equation}
(2\pi)^{3}\delta^{(3)}(0)(2 k^{0})\frac{d\rho_{ij}^{(\gamma)}(k)}{dt}=i\Bracket{[H_{\gamma g}(t),\mathcal{D}_{ij}^{(\gamma)}(k)]}
\end{equation}
Considering the Einstein Hilbert Lagrangian 
\begin{equation}
L_{g}=\frac{\sqrt{-g}}{k^{2}}g^{\alpha\lambda}(\Gamma_{\alpha\lambda}^{\beta}\Gamma^{\mu}_{\beta\mu}-\Gamma^{\mu}_{\alpha\beta}\Gamma_{\lambda\mu}^{\beta})
\end{equation}
In the weak field limit, and with the expansion $g_{\mu\nu}(x)=\eta_{\mu\nu}+k h_{\mu\nu}(x)$, expanding the Einstein Hilbert Lagrangian in $k$ and $h_{\mu\nu}(x)$ we see that Einstein's equations cancel out the linear terms in $h_{\mu\nu}(x)$ and the non-linear terms in $h_{\mu\nu}$ up to order $k$ providing us with a series of the form 
\begin{equation}
L_{g}=L_{g}^{(0)}+L_{g}^{(1)}+\mathcal{O}(k^{2})
\end{equation}
where
\begin{equation}
L_{g}^{(0)}=\frac{1}{4}\{h_{\alpha}^{\alpha,\beta}(2h^{,\lambda}_{\beta\lambda}-h_{\alpha,\beta}^{\alpha})+h^{\sigma\lambda,\alpha}(-2h_{\alpha\lambda,\sigma}+h_{\sigma\lambda,\alpha})\}
\end{equation}
describes the free propagation of gravitons and 
\begin{widetext}
\begin{equation} 
\begin{array}{c}
L_{g}^{(1)}=k\{\frac{1}{2}h^{\alpha}_{\alpha}L_{g}^{(0)}-\frac{1}{4}h^{\lambda\rho}[2 h_{\alpha,\sigma}^{\alpha}(h_{\lambda,\rho}^{\sigma}-h^{,\sigma}_{\lambda\rho})+\\
\\
+(-2 h_{\alpha\lambda,\sigma}h_{\rho}^{\sigma,\alpha}+2 h_{\sigma\lambda,\alpha}h_{\rho}^{\sigma,\alpha}+h_{\alpha\sigma,\lambda}h_{,\rho}^{\sigma\alpha})+h_{\alpha,\lambda}^{\alpha}(2 h_{\rho,\nu}^{\nu}-h_{\alpha,\rho}^{\alpha})+\\
\\
+2h_{,\nu}^{\sigma\nu}h_{\lambda\rho,\sigma}-4h_{\rho}^{\sigma,\alpha}h_{\alpha\sigma,\lambda}]\}
\end{array}
\end{equation}
\end{widetext}

The graviton field can be expanded in a Fourier series 
\begin{equation}
h_{\mu\nu}(x)=\int \frac{d^{3}q}{(2\pi)^{3}}\frac{1}{2 q^{0}}\sum_{+, \times}[b_{r}(q)h_{\mu\nu}^{(r)}e^{iqx}+b^{\dagger}_{r'}(q)h_{\mu\nu}^{(r)*}e^{-iqx}]
\end{equation}
where
\begin{equation}
[b_{r}(q),b^{\dagger}_{r'}(q')]=(2\pi)^{3}2 q^{0}\delta^{(3)}(q-q')\delta_{rr'}
\end{equation}
is the commutation relation between the graviton annihilation and creation operators and $h_{\mu\nu}^{(r)}$ are the polarisation tensors. 
The graviton propagator in the Donder gauge is given by 
\begin{equation}
D_{\mu\nu\alpha\beta}^{(g)}(q)=\frac{i}{q^{2}}(\eta_{\mu\alpha}\eta_{\nu\beta}+\eta_{\mu\beta}\eta_{\nu\alpha}-\eta_{\mu\nu}\eta_{\alpha\beta})
\end{equation}
On the electromagnetic side we also have the Maxwell Lagrangian density 
\begin{equation}
L_{\gamma}=L_{\gamma}^{(0)}+L_{\gamma}^{(1)}+L_{\gamma}^{(2)}+...
\end{equation}
with the photon propagation described by the first term
\begin{equation}
L_{\gamma}^{(0)}=-\frac{1}{16 \pi}F_{\mu\nu}F^{\mu\nu}
\end{equation}
and the other terms will give the photon graviton couplings
\begin{equation}
L_{\gamma}^{(1)}=-\frac{k}{16 \pi}(\frac{1}{2}h_{\alpha}^{\alpha}F_{\mu\nu}F^{\mu\nu}-2h^{\mu\nu}F_{\mu}^{\alpha}F_{\alpha\nu})
\end{equation}
as well as 
\begin{widetext}
\begin{equation}
L_{\gamma}^{(2)}=-\frac{k^{2}}{16\pi}\{-h_{\alpha}^{\alpha}h^{\mu\nu}F_{\mu}^{\alpha}F_{\alpha\nu}+[\frac{1}{8}(h_{\alpha}^{\alpha})^{2}-\frac{1}{4}h_{\mu\nu}h^{\mu\nu}]F_{\alpha\beta}F^{\alpha\beta}+h^{\mu\nu}h^{\alpha\beta}F_{\mu\alpha}F_{\nu\beta}+2h^{\mu\alpha}h^{\nu}_{\alpha}F_{\mu\beta}F_{\nu}^{\beta}\}
\end{equation}
\end{widetext}
These two Lagrangian densities describe each one of the vertices associated to the interaction between two photons and a graviton and respectively two photons and two gravitons. 
The second order term will only appear in higher order corrections that will not be used in the current work.
Fourier expanding the photon field one obtains
\begin{equation}
A_{\mu}(x)=\int\frac{d^{3}p}{(2\pi)^{3}}\frac{1}{2 p^{0}}\sum_{s=1,2}[a_{s}(p)\epsilon_{\mu}^{(s)}e^{ipx}+a_{s'}^{\dagger}(p)\epsilon_{\mu}^{(s)*}e^{-ipx}]
\end{equation}
with
\begin{equation}
[a_{s}(p), a_{s'}^{\dagger}(p')]=(2\pi)^{3}2 p^{0}\delta^{(3)}(p-p')\delta_{ss'}
\end{equation}
being the commutation relation between the annihilation and creation operators for photons and $\epsilon_{\mu}^{(s)}$ are the photon polarisation four-vectors where $s$ indicates the physical transversal photon polarisation. Fixing the gauge for the photon field as the Feynman gauge we obtain the propagator
\begin{equation}
D_{\mu\nu}^{(\gamma)}(p)=-4\pi \frac{i}{p^{2}}\eta_{\mu\nu}
\end{equation}
The second order S-matrix contribution to the photon-graviton scattering then is 
\begin{widetext}
\begin{equation}
S^{(2)}=-\frac{1}{2}\int d^{4}x_{1}d^{4}x_{2} T\{L_{\gamma}^{(1)}(x_{1})L_{\gamma}^{(1)}(x_{2})\}-\frac{1}{2}\int d^{4}x_{1}d^{4}x_{2}T\{L_{\gamma}^{(1)}(x_{1})L_{g}^{(1)}(x_{2})\}+i\int d^{4}x L_{\gamma}^{(2)}(x)
\end{equation}
\end{widetext}
where $T$ is the time ordering operator.
Give $p, p'$ the incoming and outgoing photon momenta and $q,q'$ the incoming and outgoing graviton momenta, we can evaluate the second order $S$-matrix. The photon-graviton Hamiltonian then turns out to be 
\begin{widetext}
\begin{equation}
H_{\gamma g}(t)=\int dq dq' dp dp' (2\pi)^{3}\delta^{(3)}(q'+p'-q-p)exp[it(q'^{0}+p'^{0}-q^{0}-p^{0})] [b^{\dagger}_{r'}(q')a^{\dagger}_{s'}(p')(M_{1}+M_{2}+M_{3})a_{s}(p)b_{r}(q)]
\end{equation}
\end{widetext}
The only contribution to the forward scattering will be that of $M_{2}$ which is

\begin{widetext}
\begin{equation}
\begin{array}{c}
M_{2}=\frac{k^{2}}{2 q\cdot q'}\{[(p\cdot p')(\epsilon\cdot \epsilon'^{*})-(\epsilon'^{*}\cdot p)(\epsilon\cdot p')][2(e\cdot q')(e'^{*})(q\cdot q')](e\cdot e'^{*})-\\
\\
-[(p\cdot p')\epsilon^{\mu}\epsilon'^{*\nu}+(\epsilon\cdot \epsilon'^{*})p^{\mu}p'^{\nu}-(p\cdot\epsilon'^{*})\epsilon^{\mu}p'^{\nu}-(\epsilon\cdot p')p^{\mu}\epsilon'^{*\nu}+(\epsilon, p \leftrightarrow \epsilon'^{*}, -p')]\times\\
\\
\times[(e\cdot e'^{*})^{2}q_{\mu}q'_{\nu}-2(e\cdot e'^{*})(q\cdot q')e_{\mu}e'^{*}_{\nu}+2(e\cdot q')(e'^{*}\cdot q)e_{\mu}e'^{*}_{\nu}+(e\cdot q')^{2}e'^{*}_{\mu}e'^{*}_{\nu}+\\
\\
+(e'^{*}\cdot q)^{2}e_{\mu}e_{\nu}-2(e\cdot e'^{*})(e\cdot q')e'^{*}{\nu}q'_{\mu}-2(e\cdot e'^{*})(e'^{*}\cdot q)e_{\nu}q_{\mu}]\}
\end{array}
\end{equation}
\end{widetext}
In order to calculate the evolution equation for the polarisation matrix of the photons in the forward scattering approximation we have to calculate the expectation value of the quantity
\begin{widetext}
\begin{equation}
\begin{array}{c}
i [H_{\gamma g}(t), \mathcal{D}_{ij}^{(\gamma)}]=

i\int dq dq' dp dp' (2\pi)^{3}\delta^{(3)}(q'+p'-q-p)M_{2}
[b_{r'}^{\dagger}b_{r}(q)a_{s'}^{\dagger}(p')a_{j}(k)2 p^{0}(2 \pi)^{3}\delta_{i s}\delta^{(3)}(p-k)-\\
\\
-b_{r'}^{\dagger}(q')b_{r}(q)a_{i}^{\dagger}(k)a_{s}(p) 2 p'^{0}(2 \pi)^{3}\delta_{j s'}\delta^{(3)}(p'-k)]
\end{array}
\end{equation}
\end{widetext}
At this point we need to use the general formula for the expectation value, namely 
\begin{equation}
\Bracket{A}=Tr(\hat{\rho}^{(i)}A)
\end{equation}
In the case of photons, given terms containing products of creation and annihilation operators, we use
\begin{equation}
\Bracket{a_{m}^{\dagger}(p')a_{n}(p)}=2 p^{0} (2\pi)^{3}\delta^{(3)}(p-p')\rho^{(\gamma)}_{mn}(p)
\end{equation}
where $\rho_{mn}^{(\gamma)}$ is the polarisation density matrix of the photons
\begin{equation}
\rho_{mn}=\frac{1}{2} 
\begin{bmatrix}
I^{(\gamma)}+Q^{(\gamma)} &   U^{(\gamma)}-iV^{(\gamma)}\\
U^{(\gamma)}+iV^{(\gamma)} &   I^{(\gamma)}-Q^{(\gamma)}\\
\end{bmatrix}
\end{equation}
We have defined the Stokes parameters as $I^{(\gamma)}$, $Q^{(\gamma)}$, $U^{(\gamma)}$, $V^{(\gamma)}$ corresponding respectively to the unpolarised intensity $I$, the linear polarisation intensities $Q$ and $U$ and the circular polarisation intensity $V$. 
For gravitons we have: 
\begin{equation}
\Bracket{b_{m}^{\dagger}(q')b_{n}(q)}=2q^{0}(2\pi)^{3}\delta^{(3)}(q-q') \rho_{mn}^{(g)}(q)
\end{equation}
producing for gravitons the polarisation density matrix 
\begin{equation}
\rho_{mn}=\frac{1}{2} 
\begin{bmatrix}
I^{(g)}+Q^{(g)} &   U^{(g)}-iV^{(g)}\\
U^{(g)}+iV^{(g)} &   I^{(g)}-Q^{(g)}\\
\end{bmatrix}
\end{equation}
which leads to the forward scattering term 
\begin{widetext}
\begin{equation}
i\Bracket{[H_{f}(t), \mathcal{D}_{ij}^{\gamma}(k)]}=i(2\pi)^{3}(0)\int dq(\delta_{is}\rho^{(\gamma)}_{s'j}(k)-\delta_{js'}\rho_{is}^{(\gamma)}(k))\rho^{(g)}_{rr'}(q)M_{2}^{r,r',s',s'}(q'=q,p=p'=k)
\end{equation}
\end{widetext}
and using the expression for $M_{2}$ we obtain 
\begin{widetext}
\begin{equation}
\begin{array}{c}
\dot{I}^{(\gamma)}=0\\
\\
\dot{Q}^{(\gamma)}=\frac{k^{2}}{k^{0}}\int dq I^{(g)}(q)(q\cdot \epsilon^{1}(k))(q\cdot \epsilon^{2}(k))V^{(\gamma)}\\
\\
\dot{U}^{(\gamma)}=-\frac{k^{2}}{2k^{0}}\int dq I^{(g)}(q)[(q\cdot\epsilon^{1}(k))^{2}-(q\cdot \epsilon^{2}(k))^{2}]V^{(\gamma)}\\
\\
\dot{V}^{(\gamma)}=\frac{k^{2}}{2k^{0}} \int dq I^{(g)}(q)[((q\cdot \epsilon^{1}(k))^{2}-(q\cdot \epsilon^{2}(k))^{2})U^{(\gamma)}-2(q\cdot \epsilon^{1}(k))(q\cdot \epsilon^{2}(k))Q^{(\gamma)}]
\end{array}
\end{equation}
\end{widetext}

Following such a scattering process on the gravitons, we can witness the entanglement of the resulting photons by means of the concurrence 
\begin{equation}
C=\sqrt{2(1-Tr(\hat{\rho}^{2}))}
\end{equation}

\section{7. photon-photon scattering mediated by gravitons}
The gravitational interaction between photons is generally assumed to be extremely weak. However, at low enough energies, the sole electromagnetic channel for photon-photon interaction, namely the creation of an electron-positron loop is exponentially disfavoured. The remaining coupling would then be gravitational, hence the exchange of virtual gravitons, which in the low energy limit would provide a larger contribution than the electron-positron pair mediated photon-photon scattering. Figure 1 presents the diagrams involving photon-photon scattering mediated by gravitons while Figure 2 presents the diagram involving the scattering of photons mediated by the virtual electron-positron loop. 
\begin{figure}
\centering
\includegraphics[scale=0.05]{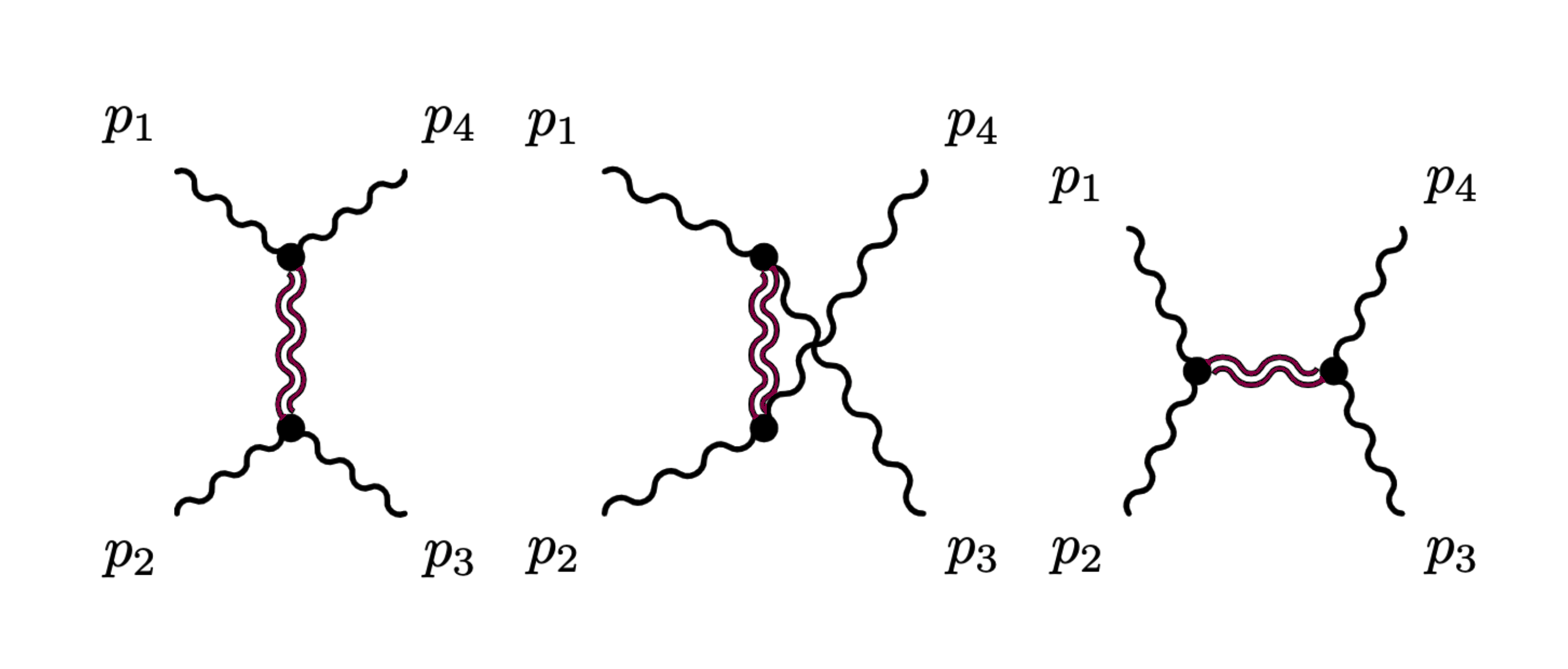}
\caption{Feynman diagrams associated to photon photon scattering mediated by gravitons}
\end{figure}

\begin{figure}
\centering
\includegraphics[scale=0.1]{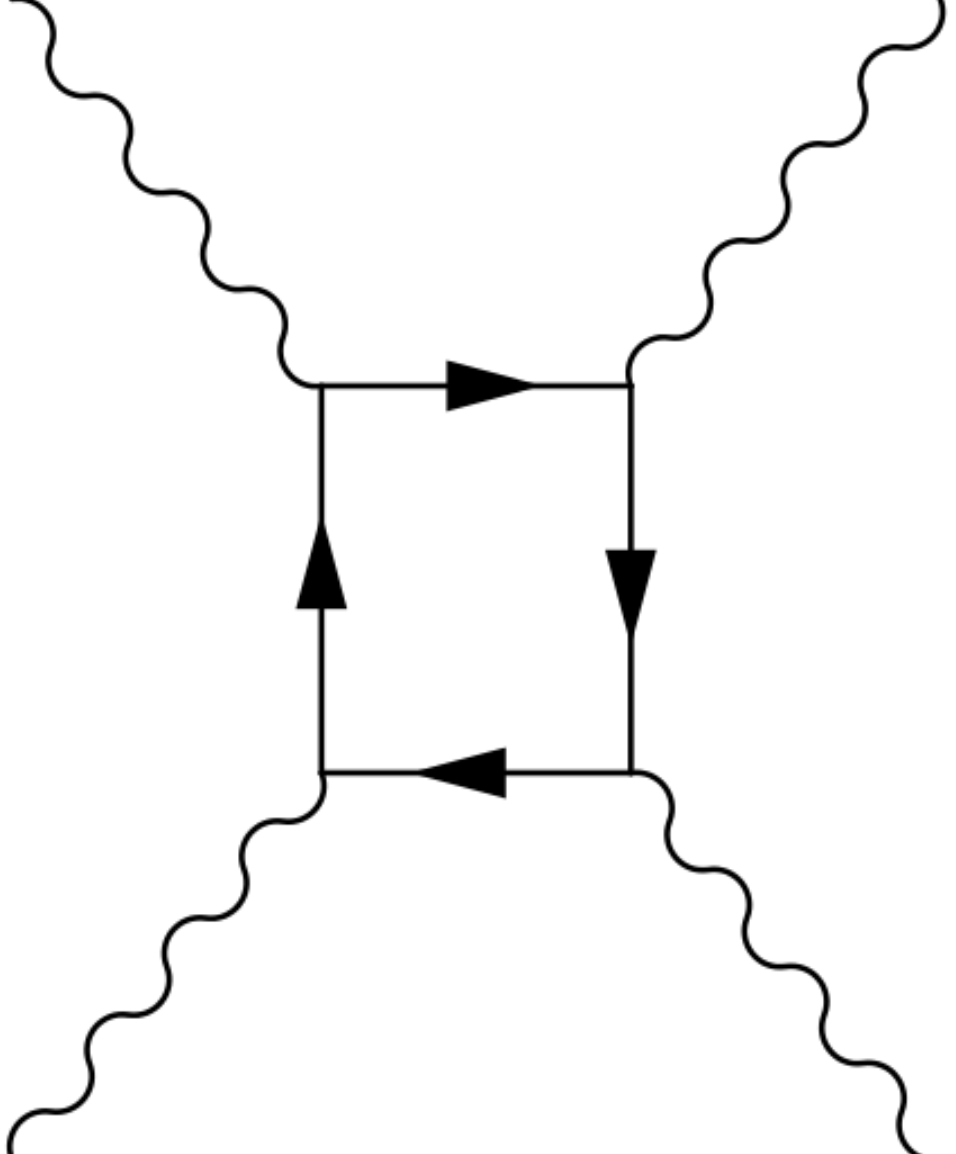}
\caption{Feynman diagrams associated to photon photon scattering mediated by a electron-positron loop}
\end{figure}

In the centre of mass system we define the propagation of the incoming photons by $p$ and $p'$ in the initial state and $q$, $q'$ in the final state. 
\begin{widetext}
\begin{equation}
\begin{array}{ccc}
p=-q, & p'=-q', & p_0=q_0=p'_0=q'_0\\
k=p'-p=-(q'-q), & s=p'+p=-(q'+q), &
\end{array}
\end{equation}
\end{widetext}
The scattering process in which the first initial photon $p$ emits a graviton ending up in the final state $p'$ and the second initial photon is the receiver of the graviton having the initial state $q$ and ends up in the final state $q'$ is described by the scattering matrix element 
\begin{widetext}
\begin{equation}
\begin{array}{c}
M_{1}=i (2\pi)^{4}\delta(p+q-p'-q') a_{i}^{\dagger}(p')a_{j}(p)a_{l}^{\dagger}(q')a_{m}(q)(\frac{k^{2}c\hbar}{8 p_{0}^{4}})\times \\
\\
\times [2p^{2}_{0}(\delta_{il} k_{j} k_{m}+ \delta_{jm} k_{i} k_{l})-\\
\\
-(2p_{0}^{4} +p_{0}^{2} k^{2} -\frac{1}{4} k^{4} )\delta_{il} \delta_{jm}+\\
\\
+2p_{0}^{4}(\delta_{ij} \delta_{lm} +\delta_{im}\delta_{jl})]\\
\end{array}
\end{equation}
\end{widetext}
and for the associated rotated diagram we have
\begin{widetext}
\begin{equation}
\begin{array}{c}
M_{2}=i (2\pi)^{4}\delta(p+q-p'-q') a_{i}^{\dagger}(p')a_{j}(p)a_{l}^{\dagger}(q')a_{m}(q)(\frac{k^{2}C\hbar}{8p_{0}^{4}})\times\\
\\
(2p_{0}^{2}(\delta_{il}k_{j}k_{m}+\delta_{jm}k_{i}k_{l})-\\
\\
-(2p_{0}^{4}+p^{2}_{0}k^{2}-\frac{1}{4}k^{4})\delta_{il}\delta_{jm}+\\
\\+2p_{0}^{4}(\delta_{ij}\delta_{lm}+\delta_{im}\delta_{jl}))
\end{array}
\end{equation}
\end{widetext}
We may choose one of the polarisation vectors of each photon in the plane of scattering and the other polarisation vector perpendicular to it, leading to a transformation towards circularly polarised photons. We then add up the contributions of the two scattering matrices giving 
\begin{equation}
\begin{array}{c}
M_{1}+M_{2}=(-i/c\hbar)(2\pi)^{4}\delta(p+q-p'-q)\times\\
\times [V_{++}(k)a_{+}^{\dagger}(p')a_{+}^{\dagger}(q')a_{+}(q) + \\
+ V_{--}(k)a_{-}^{\dagger}(p')a_{-}(p)a_{-}^{\dagger}(q')a_{-}(q)+\\
+ v_{+-}(k)a_{+}^{\dagger}(p')a_{+}(p)a_{-}^{\dagger}(q')a_{-}(q)+\\
+v_{-+}(k)a_{-}^{\dagger}(p')a_{-}(p)a_{+}^{\dagger}(q')a_{+}(q)]
\end{array}
\end{equation}
where
\begin{equation}
\begin{array}{c}
V_{++}(k)=V_{--}(k)=-(2c^{2}\hbar^{2}p_{0}^{2}\kappa^{2}/k^{2}),\\
V_{+-}(k)=V_{-+}(k)=-(2c^{2}\hbar^{2}p_{0}^{2}\kappa^{2}/k^{2})\times \\
\times [1-(k^{2}/4p_{0}^{2})-(k^{4}/16p_{0}^{4})+(k^{6}/64p_{0}^{6})]\\
\end{array}
\end{equation}
The creation and annihilation operators are defined according to the photons they create or annihilate as being parallel to their direction of motion (subindex plus), or antiparallel (subindex minus). 
\section{8. Quantum Boltzmann equation for graviton mediated photon-photon scattering}
To determine the Stokes parameters that define the polarisation state of the photons in our scattering process, we need to write the quantum Boltzmann equation in terms of the polarisation density matrix and the associated interaction Hamiltonian. 
\begin{widetext}
\begin{equation}
(2\pi)^{3}\delta^{(3)}(0)(2k^{0})\frac{d\rho_{ij}(k)}{dt}=i\Bracket{[H_{I}(0),\mathcal{D}_{ij}(k)]}-\frac{1}{2}\int_{-\infty}^{\infty} dt \Bracket{[H_{I}(t), [H_{I}(0),\mathcal{D}_{ij}(k)]]}
\end{equation}
\end{widetext}
$H_{I}$ is the effective interaction hamiltonian which will have to describe the coupling of the photons to gravitons, and $\mathcal{D}_{ij}$ is the photon number operator. 
The expectation value of a general operator $A$ can be defined as 
\begin{equation}
\Bracket{A(k)}=Tr[\rho A(k)]=\int \frac{d^{3}p}{(2\pi)^{3}}\Bracket{p|\rho A(k)|p}
\end{equation}
where the density operator is obtained via the integral of the number operator weighted by the density matrix
\begin{equation}
\rho = \int\frac{d^{3}p}{(2\pi)^{3}}\rho_{ij}(p)\mathcal{D}_{ij}(p)
\end{equation}
The forward scattering term, the first term in Boltzmann's equation is the one that will generate the couplings between the different polarisation states. 
The interaction hamiltonian can be generally written as
\begin{widetext}
\begin{equation}
H_{I}=\int dp^{1}dp^{2}dp^{3}dp^{4}(2\pi)^{3}\delta^{(3)}(p^{3}+p^{4}-p^{1}-p^{2})\times 3 M(p^{1}, r; p^{2}, s; p^{3}, r'; p^{4}, s')a_{r'}^{\dagger}(p^{3}) a_{r}(p^{1})a_{s'}^{\dagger}(p^{4})a_{s}(p^{2})
\end{equation}
\end{widetext}
where we defined $M$ to be the amplitude of this interaction depending on the photon momenta and the photon polarisation indices $r,r',s,s' =1,2$. $1, 2$ represent the independent transverse photon polarisations. 
Now we can determine the forward scatter term from the quantum Boltzmann equation as
\begin{widetext}
\begin{equation}
\begin{array}{c}
\Bracket{[H_{I}(0), D_{ij}(k)]}=\int dp^{1}dp^{2}dp^{3}dp^{4}(2\pi)^{(3)}(p^{3}+p^{4}-p^{1}-p^{2})\times 3 M(p^{1}, r; p^{2}, s; p^{3}, r'; p^{4}, s')\\
 \Bracket{a_{r'}^{\dagger}(p^{3})a_{r}(p^{1})a_{s'}^{\dagger}(p^{4})a_{s}(p^{2})a_{i}^{\dagger}(k)a_{j}(k)-a_{i}^{\dagger}(k)a_{j}(k)a_{r'}^{\dagger}(p^{3})a_{r}(p^{1})a_{s'}^{\dagger}(p^{4})a_{s}(p^{2})}
\end{array}
\end{equation}
\end{widetext}
The expectation value of the photon number operator calculated as a product of photon creation and annihilation operator is given by
\begin{equation}
\Bracket{a_{m}^{\dagger}(p)a_{n}(p')}=(2\pi)^{3}2p^{0}\delta^{(3)}(p-p')\rho_{mn}(p)
\end{equation}
Using Wick's theorem for the forward scattering term we obtain the quantum Boltzmann equation as
\begin{widetext}
\begin{equation}
\begin{array}{c}
\frac{d\rho_{ij}^{\gamma}(k)}{dt}=\frac{3i}{2k^{0}}\int dp ([\delta_{is}\delta_{rs'}\rho^{\gamma}_{r'j}(k)-\delta_{jr'}\delta_{rs'}\rho_{is}^{\gamma}(k)+\delta_{jr'}\rho_{is}^{\gamma}(k)\rho^{b}_{s'r}(p)-\delta_{is}\rho_{r'j}(k)\rho_{s'r}^{b}(p)]M(p,r;k,s;k,r';p,s')+\\
(\delta_{is}\rho_{s'j}^{\gamma}(k)\rho_{r'r}^{b}(p)-\delta_{s'j}\rho_{is}^{\gamma}(k)\rho_{r'r}^{b}(p))M(p,r;k,s;p,r';k,s')+\\
+[\delta_{ir}\rho^{\gamma}_{s'j}(k)\rho_{r',s}^{b}(p)-\delta_{js'}\rho_{ir}^{\gamma}(k)\rho_{r's}^{b}(p)]
M(k,r;p,s;p,r';k,s')+\\
+[\delta_{ir}\rho_{r'j}^{\gamma}(k)\rho_{s's}^{b}(p)-\delta_{jr'}\rho_{ir}^{\gamma}(k)\rho_{s's}^{b}(p)]
M(k,r;p,s;k,r';p,s'))
\end{array}
\end{equation}
\end{widetext}

The photon amplitude can be written in a general form, such that it has a universal representation for the scattering of photons on photons that only depends parametrically on the interaction hamiltonian. After this form has been written, the interaction Hamiltonian for the gravitational coupling between photons can be introduced for the final result. 
The photon-photon scattering amplitude can be parametrised as 
\begin{equation}
M=M_{\mu\nu\lambda\sigma}(1234)\epsilon_{\mu}^{1}\epsilon_{\nu}^{2}\epsilon_{\lambda}^{3}\epsilon_{\sigma}^{4}
\end{equation}
with the obvious notation for the photon polarisation and momenta. Gauge symmetry implies the obvious zero identities between the momenta and the amplitudes while the crossing symmetry provides the form of the amplitude as a sum over all $4!$ permutations of the external photons. 
Then we can re-write the scattering amplitude as a tensorial expansion in terms of rank four independent tensors and the coefficients are scalar amplitudes that define the universal kinematics and depend on the Mandelstam variables
\begin{equation}
M_{\mu\nu\lambda\sigma}(1234)=\sum_{i=1}^{5}G_{i}(s,t,u)\cdot T^{(i)}_{\mu\nu\lambda\sigma}(1,2,3,4)
\end{equation}
With this expression in absolutly general terms we can write 
\begin{widetext}
\begin{equation}
\begin{array}{l}
T_{\mu\nu\lambda\sigma}^{(1)}(1234)=f^{(1)}_{\mu\nu\lambda\sigma}(1234),\\
T_{\mu\nu\lambda\sigma}^{(2)}(1234)=f^{(1)}_{\lambda\nu\mu\sigma}(3214),\\
T_{\mu\nu\lambda\sigma}^{(3)}(1234)=f^{(1)}_{\sigma\nu\lambda\mu}(4231),\\
T_{\mu\nu\lambda\sigma}^{(4)}(1234)=f^{(2)}_{\mu\nu\lambda\sigma}(1234)+f^{(2)}_{\mu\nu\sigma\lambda}(1243)+f^{(2)}_{\nu\lambda\mu\sigma}(2314),\\
T_{\mu\nu\lambda\sigma}^{(5)}(1234)=f^{(3)}_{\mu\nu\lambda\sigma}(1234)+f^{(3)}_{\nu\mu\sigma\lambda}(2143)+f^{(3)}_{\lambda\sigma\mu\nu}(3412)+f^{(3)}_{\sigma\lambda\nu\mu}(4312)+\\

\qquad\qquad\qquad\quad +f^{(3)}_{\mu\lambda\nu\sigma}(1324)+f^{(3)}_{\lambda\mu\sigma\nu}(3142)+f^{(3)}_{\nu\sigma\mu\lambda}(2413)+f^{(3)}_{\sigma\nu\lambda\mu}(4231)+\\

\qquad\qquad\qquad\quad +f^{(3)}_{\mu\sigma\lambda\nu}(1432)+f^{(3)}_{\sigma\mu\nu\lambda}(4123)+f^{(3)}_{\lambda\nu\mu\sigma}(3214)+f^{(3)}_{\nu\lambda\sigma\mu}(2341)\\
\end{array}
\end{equation}
\end{widetext}
with the tensor basis depending only on the metric tensor $g_{\mu\nu}$, the photon four-momenta, and the polarisation vectors

\begin{equation}
\begin{array}{l}
f_{\mu\nu\lambda\sigma}^{(1)}(1234)=p_{\mu}^{2}p_{\nu}^{1}p_{\lambda}^{4}p_{\sigma}^{3}-\\
\qquad\qquad\qquad\quad  -(p^{3}\cdot p^{4})g_{\lambda\sigma}p_{\mu}^{2}p_{\nu}^{1}-\\
\qquad\qquad\qquad\quad -(p^{1}\cdot p^{2})g_{\mu\nu}p_{\lambda}^{4}p_{\sigma}^{3}+\\
\qquad\qquad\qquad\quad +(p^{1}\cdot p^{2})(p^{3}\cdot p^{4})g_{\mu\nu}g_{\lambda\sigma},\\
\\
\end{array}
\end{equation}

\begin{equation}
\begin{array}{l}
f^{(2)}_{\mu\nu\lambda\sigma}(1234)=p_{\mu}^{2}p_{\nu}^{3}p_{\lambda}^{4}p_{\sigma}^{1}+p_{\mu}^{4}p_{\nu}^{1}p_{\lambda}^{2}p_{\sigma}^{3}-\\
\qquad\qquad\qquad\quad   -(p^{1}\cdot p^{4})g_{\lambda\sigma}p_{\mu}^{2}p_{\nu}^{3}-\\
\qquad\qquad\qquad\quad  -(p^{3}\cdot p^{4})g_{\nu\lambda}p_{\nu}^{3}p_{\lambda}^{4}+\\
\qquad\qquad\qquad\quad  +(p^{1}\cdot p^{2})g_{\lambda\sigma}p_{\nu}^{3}p_{\mu}^{4}-\\
\qquad\qquad\qquad\quad  -(p^{1}\cdot p^{2})g_{\nu\lambda}p_{\mu}^{4}p_{\sigma}^{3}+\\
\qquad\qquad\qquad\quad  +(p^{2}\cdot p^{3})g_{\mu\sigma}p_{\nu}^{1}p_{\lambda}^{4}-\\
\qquad\qquad\qquad\quad  -(p^{2}\cdot p^{3})g_{\lambda\sigma}p_{\mu}^{4}p_{\nu}^{1}-\\
\qquad\qquad\qquad\quad  -(p^{3}\cdot p^{4})g_{\mu\sigma}p_{\nu}^{1}p_{\lambda}^{2}+\\
\qquad\qquad\qquad\quad  +(p^{1}\cdot p^{4})(p^{2}\cdot p^{3})g_{\mu\nu}g_{\lambda\sigma}+\\
\qquad\qquad\qquad\quad  +(p^{1}\cdot p^{2})(p^{3}\cdot p^{4})g_{\mu\sigma}g_{\nu\lambda},\\
\\
\end{array}
\end{equation}

\begin{equation}
\begin{array}{l}
f^{(3)}_{\mu\nu\lambda\sigma}(1234)=(p^{3}\cdot p^{4})p^{2}_{\mu}p_{\nu}^{1}p_{\lambda}^{1}p_{\sigma}^{1}-\\
\qquad\qquad\qquad\quad  -(p^{1}\cdot p^{3})p_{\mu}^{2}p_{\nu}^{1}p_{\lambda}^{4}p_{\sigma}^{1}-\\
\qquad\qquad\qquad\quad  -(p^{1}\cdot p^{4})p_{\mu}^{2}p_{\nu}^{1}p_{\lambda}^{1}p_{\sigma}^{3}+\\
\qquad\qquad\qquad\quad  +(p^{1}\cdot p^{3})(p^{1}\cdot p^{4})g_{\lambda\sigma}p^{2}_{\mu}p_{\nu}^{1}+\\
\qquad\qquad\qquad\quad  +(p^{2}\cdot p^{3})(p^{1}\cdot p^{4})g_{\mu\sigma}p^{4}_{\lambda}p_{\sigma}^{1}-\\
\qquad\qquad\qquad\quad  -(p^{1}\cdot p^{2})(p^{3}\cdot p^{4})g_{\mu\nu}p^{1}_{\lambda}p_{\sigma}^{1}+\\
\qquad\qquad\qquad\quad  +(p^{1}\cdot p^{2})(p^{1}\cdot p^{4})g_{\mu\nu}p^{1}_{\lambda}p_{\sigma}^{3}-\\
\qquad\qquad\qquad\quad  -(p^{1}\cdot p^{2})(p^{1}\cdot p^{3})(p^{1}\cdot p^{4})g_{\mu\nu}g_{\lambda\sigma}\\
\end{array}
\end{equation}

For the gravitational interaction between the two photons we use the expansion 
\begin{equation}
\begin{array}{l}
T_{\mu\nu\beta\alpha}(p',p)=p'_{\alpha}(p_{\mu}\eta_{\beta\nu}+p_{\nu}\eta_{\beta\mu})+\\
\qquad\qquad\qquad\quad  +p_{\beta}(p'_{\mu}\eta_{\alpha\nu}+p'_{\nu}\eta_{\alpha\mu})-\\
\qquad\qquad\qquad\quad  -\eta_{\alpha\beta}(p'_{\mu}p_{\nu}+p_{\mu}p'_{\nu})+\\
\qquad\qquad\qquad\quad  +\eta_{\mu\nu}((p'\cdot p)\eta_{\alpha\beta}-p_{\beta}p'_{\alpha})-\\
\qquad\qquad\qquad\quad  -p'\cdot p(\eta_{\mu\alpha}\eta_{\nu\beta}+\eta_{\mu\beta}\eta_{\nu\alpha})
\end{array}
\end{equation}
The three diagrams included will correspond again to the two diagrams mentioned above and the crossing diagram. 
The internal graviton line is associated with the integration over the graviton momentum $q$ and graviton propagator 
\begin{equation}
\frac{i \mathcal{P}}{q^{2}+i\epsilon}=\frac{i}{2(q^{2}+i\epsilon)}(\eta_{\mu\alpha}\eta_{\nu\beta}+\eta_{\mu\beta}\eta_{\nu\alpha}-\eta_{\mu\nu}\eta_{\alpha\beta})
\end{equation}
The resulting amplitude is then 

\begin{equation}
\begin{array}{l}
M_{a}=(\epsilon^{\beta}(p_{3}))^{*}\epsilon^{\alpha}(p_{1})T_{\mu\nu\beta\alpha}(p_{3},p_{1})\\
\frac{\mathcal{P}^{\mu\nu\rho\sigma}}{(p_{1}-p_{2})^{2}}(\epsilon^{\delta}(p_{4}))^{*}\epsilon^{\gamma}(p_{1})T_{\rho\sigma\delta\gamma}(p_{4}, p_{2})
\end{array}
\end{equation}

The summation of the three tree diagrams will then give the total amplitude. 

\begin{figure}
\centering
\includegraphics[scale=0.5]{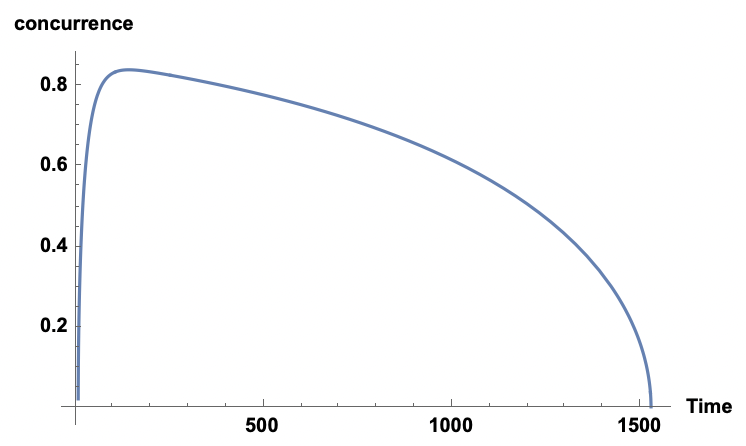}
\caption{Evolution of entanglement with time as witnessed by concurrence (equation 169)}
\end{figure}

\section{9. Polarisation density matrix and entanglement}
Now that we obtained general forms for the amplitudes and are able to calculate the polarisation density matrix, the next step is to use the form of this density matrix to determine the amount of entanglement. The basic idea is that through the process of graviton exchange that mediates the interaction between the photons, entanglement is produced. 

Therefore evolving the polarisation density matrix by means of Boltzmann's equation we obtain a mixing state of the photon polarisation, the presence of entanglement being determined by means of the usual indicators. 
In particular, Boltzmann type equations are suitable for the analysis of open systems and in particular of systems of many particles. The discussion of this paper differs from the usual approach in which entanglement is calculated starting from a single interaction defined by a specific process and therefore can be applied to more complex situations like interactions between photons that appear as a photon gas in the early universe and the interactions between them being mediated by early cosmological gravitons. Because of this, using master equations of the type that allows the analysis of open systems is particularly useful in discussing situations like the entangling effect of early cosmological expansion, inflation, and other processes involving early universe photons. 

As such, a numerical analysis has been performed involving both contributions to the Boltzmann equation, and the results are being presented and analysed in what follows. 
We numerically solve this equation for a density matrix depending explicitly on time, and obtain the matrix elements as they evolve during the interaction. With this we can probe the evolution of entanglement of the system of polarised photons after the interaction and hence detect the generation of entanglement between the two photons as mediated by the graviton. The density matrix contains $I^{(\gamma)}$ as the parameter encoding the intensity of the unpolarised radiation, $Q^{(\gamma)}$ and $U^{(\gamma)}$ defining the intensity of the linear polarisation, while $V^{(\gamma)}$ being the circular polarisation. With these, using concurrence as the entanglement witness we obtain the time evolution of entanglement in the considered process. The results, after considering the graviton as a mediator of interaction, describe two regions: the first region, in the timeframe of the interaction, the two pairs are being entangled and we notice an increase in the entanglement concurrence up to a maximal value of $0.8403$, in a relative short period of time. The value of the entanglement then is following a slowly decreasing plateau, in which we suppose that the entanglement is slowly dissipating in the photon gas. This results from the numerical solution of the Boltzmann equation which results in the density matrix then used in the concurrence formula that measures entanglement. The evolution of the entanglement is propagated over a time up to three orders of magnitude larger than the time required to generate the entanglement. I must underline that the solution to the equations were obtained numerically in Mathematica and that they are trustworthy within a few factors beyond the regime in which the graviton mediates the interaction and produces entanglement. In extreme cases in which we analyse the evolution at larger times, the entanglement indeed decreases, as expected, but the evolution in the final stages may not be perfectly captured due to numerical issues. Indeed, as presented in ref. [31] it is possible, and in fact likely that the final decay of entanglement to be exponential in nature, (or eventually algebraic given a high number of participating particles [31]). The method itself employed in this article however can make predictions about the large timeframe evolution of entanglement, and I intend to study this aspect in a future article. A more detailed implementation of the mechanisms of decoherence will then be required.
The mechanism through which entanglement is spread into the environment is captured by the quantum Boltzmann equation.
The dependence of the concurrence on time is shown in figure 3.
The exchange of gravitons rapidly generates entanglement between the photons, which starts decaying in the environment over a timescale larger than the interaction timescale. The process through which this happens is described by the usual techniques of decoherence. Of course, the process is fairly quick nevertheless, but given that we are discussing about a photon gas in cosmological settings (or even in laboratory settings) the timescale of decoherence will be more extended than the timescale over which entanglement is created. The process of decoherence occurs when the quantum system that manifests some internal quantum properties (superposition, entanglement, etc.) becomes entangled with its environment. As we describe a gas of photons subject to gravitational interactions, the evolution involves two effects: the entangling effect of the gravitational interaction and at the same time the decoherence manifested by the entangling of our reference photon with the gas of particles it is travelling through, which is the fundamental model for the quantum Boltzmann equation. 
\begin{widetext}
\begin{figure}
\centering
\begin{tabular}{c c}
\includegraphics[width=.95\linewidth,height=95pt]{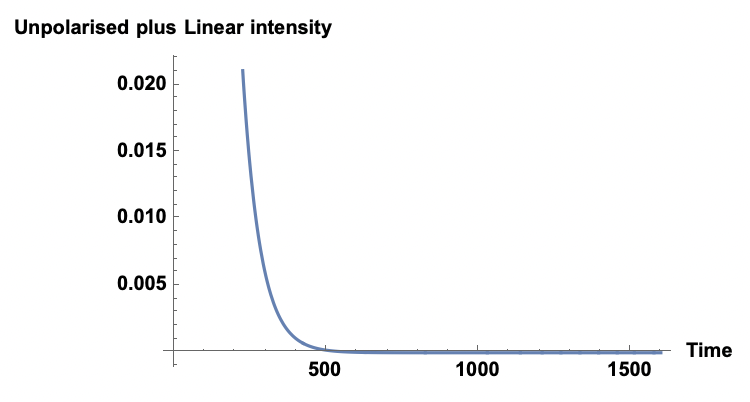} & \includegraphics[width=.95\linewidth,height=95pt]{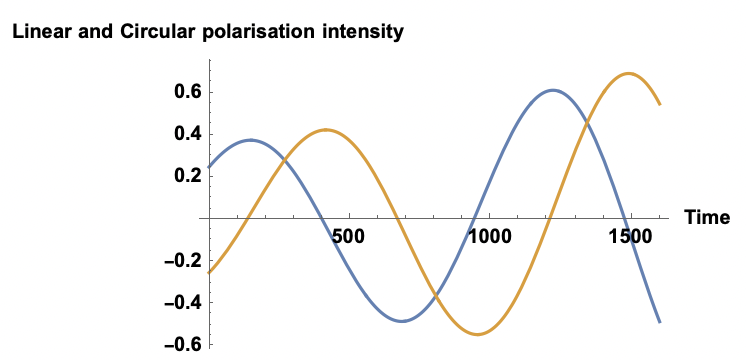} 
\end{tabular}
\vspace{\floatsep}
\begin{tabular}{c c}
\includegraphics[width=.95\linewidth,height=95pt]{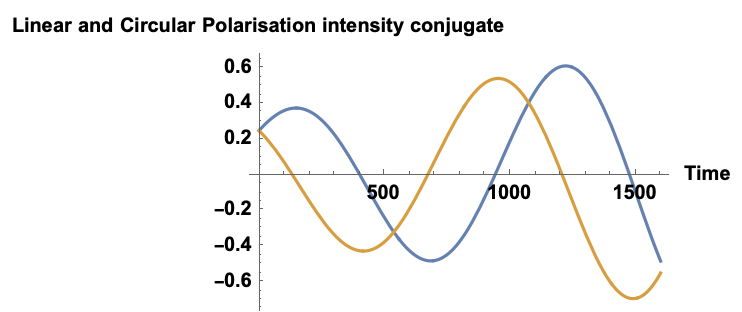} & \includegraphics[width=.95\linewidth,height=95pt]{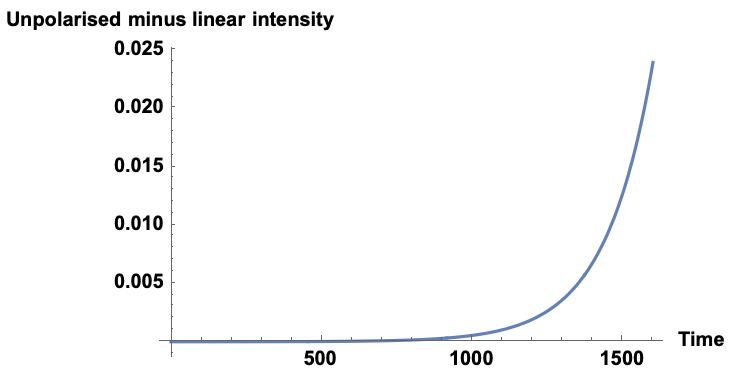}
\end{tabular}
\caption{Evolution of polarisation density matrix elements; density matrix elements corresponding to the respective polarisation states defined by means of Stokes parameters (see eq. 164, 166)}
\end{figure}
\end{widetext}

As we used a method based on a master equation, in our case, the quantum Boltzmann master equation for a density matrix describing polarisation intensities, we expected to obtain a description of the evolution of entanglement in an open system, and hence we were able to trace the evolution of entanglement as it is spreading in the environment. The evolution of the various polarisation terms is being shown in figure 4. 

With this analysis we can establish therefore several cosmological periods according to the generation of entanglement between photons. We would have an early period in which structure in the universe would present high entanglement mediated by gravitons, followed by a slow decay over a time frame of 3 orders of magnitude. 

The evolution equation for the entanglement witness we used is shown in Figure 3, while the density matrix elements corresponding to the respective polarisation states defined by means of Stokes parameters are shown in figure 4. 

\section{10. conclusion}

The question of whether the graviton is a quantum system or not is answered in the positive in this article, with the calculations showing that the graviton mediated interaction does indeed generate entangled pairs and therefore the graviton cannot behave classically. Moreover, the procedure used to obtain this result is based on the Quantum Boltzmann equation, an equation dealing with systems far from equilibrium and in particular with open systems. By this approach we can study the evolution of entanglement in a photon gas, subject to gravitational interactions. This is particularly amenable to the early stages of the universe when the photon density was high and gravitational interaction was relevant for the coupling between photons. The resulting entanglement would have spread over large distances resulting in correlations that would abruptly stop leading to a decrease in correlation at finer grained scales (say, at the level of galaxies or star systems), while keeping an overall correlations at larger cosmological scales. Moreover, this is by far not the only process that can be analysed in this form. Indeed, it is possible to analyse the interaction between a photon gas and a transiting gravitational wave, and at the same time to study the entanglement produced between those objects. This would lead to understanding larger scale correlations as well as to a new type of probing of the early universe, by means of entanglement. Among the processes that can be analysed and for which entanglement dynamics can be obtained are those involving helicity coupling of photons to massive bodies, Chern-Simons theory entanglement, etc. The most important aspect however is that we can, using this approach, analyse the dynamics of entanglement as evolving in time, and hence we can predict how entanglement should behave in an open system as well as in systems that are far from equilibrium. The use of master equations to grasp quantum phenomena is notoriously difficult, as the full quantum master equation usually involves extremely hard computations. However, due to the development of quantum master equations that can reflect most quantum properties, and finally due to the development in the field of quantum Boltzmann equations, this new field of research is within our reach. 
\section{acknowledgement}
I wish to thank A. Mazumdar and S. Bose for some constructive remarks regarding some parts of the calculations performed in this article. 
\section{Data Availability Statement}
This article has not generated any new data.

\end{document}